\documentclass[a4paper,11pt]{article}
\usepackage{jinstpub} % for details on the use of the package, please see the JINST-author-manual
\usepackage{subcaption}
\usepackage{graphicx}
\usepackage{lineno}
\title{\boldmath A compact beta particle momentum detector}

\author[a,1]{D. Kodroff,\note{Corresponding author.}}
\author[a]{M. Hu,}
\author[a,b]{K. Adcock,}
\author[a]{R. Carney,}
\author[a]{P. Denes,}
\author[a]{M. Garcia-Sciveres,}
\author[a]{A. Goldschmidt,}
\author[a,b]{V. Gomez,}
\author[a]{P. Sorensen,}
\author[a]{T. Villabona,}
\author[a,b,2]{D. Carney,\note{Corresponding author.}}

\affiliation[a]{Lawrence Berkeley National Laboratory, Berkeley, CA 94720-8099, USA}
\affiliation[b]{University of California, Berkeley, CA 94720-8099, USA}

\emailAdd{danielkodroff@lbl.gov}
\emailAdd{carney@lbl.gov}

\abstract{We present the design, development, and first calibration results of a compact beta electron detector for the Quantum Invisible Particle Sensor (QuIPS) experiment. The QuIPS electron detector is designed to reconstruct the full momentum vector of $\beta$ particles emitted from radioisotope-doped optically levitated nanospheres in ultra-high vacuum (UHV), enabling a measurement of the neutrino momentum and a search for heavy sterile neutrinos. The detector comprises two thinned CMOS detectors for directional tracking and a plastic scintillator read out by silicon photomultipliers (SiPMs) for calorimetry. The entire assembly must operate inside an existing optical trapping vacuum chamber at pressures below $10^{-7}$~mbar, imposing stringent constraints on material selection, power dissipation, outgassing, and compactness. We demonstrate sensitivity to $\beta$-decay electrons with energies as low as 100~keV and a detection efficiency of 35\% above 500~keV, the primary window of interest for a heavy sterile neutrino search. The CMOS tracker resolves the momentum direction at the mrad scale and reconstructs the $\beta$ emission vertex with sub-mm precision, while the scintillator-SiPM system achieves an energy resolution of 5\% at 1~MeV.}


\begin{document}
\maketitle
\flushbottom

\section{Introduction}
\label{sec:intro}

The Quantum Invisible Particle Sensor (QuIPS) experiment, first conceptualized in Ref.~\cite{Carney:2022pku}, represents a new approach to neutrino measurements in nuclear $\beta$ decays. Unlike conventional methods, where the signal is typically inferred from distortions in the $\beta$ energy spectrum, QuIPS instead aims at direct kinematic reconstruction of the neutrino. The concept employs a levitated optomechanical sensor: dielectric silica nanospheres of order 100~nm diameter with femtogram-scale masses, optically trapped and feedback-cooled to their quantum ground state of center-of-mass motion in ultra-high vacuum (UHV)~\cite{Delic:2019xqd,Tebbenjohanns:2021tgw,Magrini2021}. These systems have been demonstrated to be ultrasensitive force and impulse sensors~\cite{Ranjit:2016dlq,Monteiro:2020qiz,Tseng:2025rlo,Tseng:2026flh}. The QuIPS concept builds upon this by embedding the silica nanospheres with $\beta$-emitting radioisotopes. When a $\beta$-decay occurs within an optically trapped nanosphere, the emitted $\beta$ and neutrino will escape the sphere, imparting a measurable momentum impulse to the center-of-mass of the sphere. The impulse is read out via the optical detection system, and a surrounding particle detector measures the momentum of the outgoing $\beta$ particle. Conservation of momentum on the sphere's center of mass then allows inference of the neutrino momentum on an event-by-event basis. This kinematic reconstruction provides a probe of neutrino mass and enables searches for additional heavy neutrino mass eigenstates. Such a state, like a heavy sterile neutrino, if it exists, could be produced through mixing with the electron neutrino during the decay and carry away anomalously large momentum, producing a distinctive signature in the spectra of reconstructed neutrino momentum and $\beta$ energy.

The realization of this physics program places stringent demands on the particle detector surrounding the nanosphere. To match the precision of the optomechanical recoil measurement, the detector must reconstruct the full momentum vector of each emitted $\beta$ particle, requiring precise directional tracking, accurate energy measurement, and high detection efficiency. The experiment will initially study $\beta$-decays of $^{90}$Y (Q-value 2.279~MeV). Therefore, the detector must fully contain electrons from keV energies up to this endpoint, since any undetected energy loss biases the inferred neutrino momentum and can mimic a heavy sterile neutrino signature. The tracker must in addition vertex the $\beta$ trajectory to the point-like nanosphere origin to better than 1~mm, in order to reject backgrounds such as electrons scattered from the trapping optics and cosmic rays. Finally, the entire assembly must operate within the optical trapping chamber at pressures below 10$^{-7}$~mbar, imposing constraints on material selection, power dissipation, outgassing, and compactness. It must further avoid interfering with the trapping laser or optomechanical readout, and withstand excess scattered laser power.

In this paper we present the design, construction, and first calibration results of a detector that meets these requirements. The system combines two thinned CMOS sensors for directional $\beta$ tracking with an EJ-200 plastic scintillator read out by silicon photomultipliers (SiPMs) for calorimetry, all operating in UHV. Through a data-driven calibration, we demonstrate an overall detection efficiency of 35\% for electrons above 500~keV, an energy resolution of 5\% at 1~MeV with a linear response to the $^{90}$Y endpoint, and vertex reconstruction of the $\beta$ emission point to better than 1~mm. We further show that this pointing capability cleanly discriminates collimated $\beta$ particles from isotropically incident cosmic rays, validating the background-rejection strategy for the deployed experiment. To our knowledge, this constitutes the first integrated electron tracking and calorimetry detector developed for operation in concert with an optically levitated nanosphere.

\section{Design}
\label{sec:design}

The challenge of detecting and tracking electrons emitted in nuclear $\beta$ decays has arisen across several areas of precision physics, and a variety of detector technologies have been developed to reconstruct their energy and, in some cases, their direction. Experiments primarily targeting the $\beta$ energy spectrum, such as those measuring the tritium endpoint~\cite{KATRIN:2024cdt,Project8:2017nal}, the $^{187}$Re endpoint~\cite{ANDREOTTI2007208}, or mapping spectral shapes with cryogenic microcalorimeters~\cite{ACCESS:2023gdy}, achieve excellent energy resolution but do not reconstruct the electron direction. Directional and momentum information is instead accessed by a smaller class of detectors. Precision neutron $\beta$-decay experiments~\cite{Nab:2018toa,Hassan:2020hrj} and double-$\beta$ detectors, such as high-pressure time-projection chambers~\cite{NEXT:2023daz} and segmented calorimeters~\cite{SuperNEMO:2010wnd}, reconstruct electron trajectories to measure $\beta$-decay correlations and event topologies, respectively, from tens of keV up to the MeV regime. These approaches, however, generally measure ensemble correlations or reconstruct direction with limited resolution. By contrast, the requirement of this experiment is to reconstruct the full momentum vector of individual electrons on an event-by-event basis, in strict coincidence with a recoil measurement of the source itself, enabling direct kinematic reconstruction of the emitted neutrino. Historically, this capability has been limited to Paul and Penning traps~\cite{Delahaye:2018nok,SHIDLING2021116636,Burkey:2022gpb} and magneto-optical traps~\cite{Vetter:2008zz,Muller:2022jew}. Similar platforms have also been explored to perform dedicated keV-scale sterile neutrino searches~\cite{smith2019proposed,martoff2021hunter}. The QuIPS approach extends this event-by-event coincidence technique to a solid-state, optically levitated source, replacing the trapped ion or atom with a radioisotope-doped nanosphere whose recoil is read out optomechanically.

\begin{figure}[!t]
	\centering
    \includegraphics[width=0.84\columnwidth]{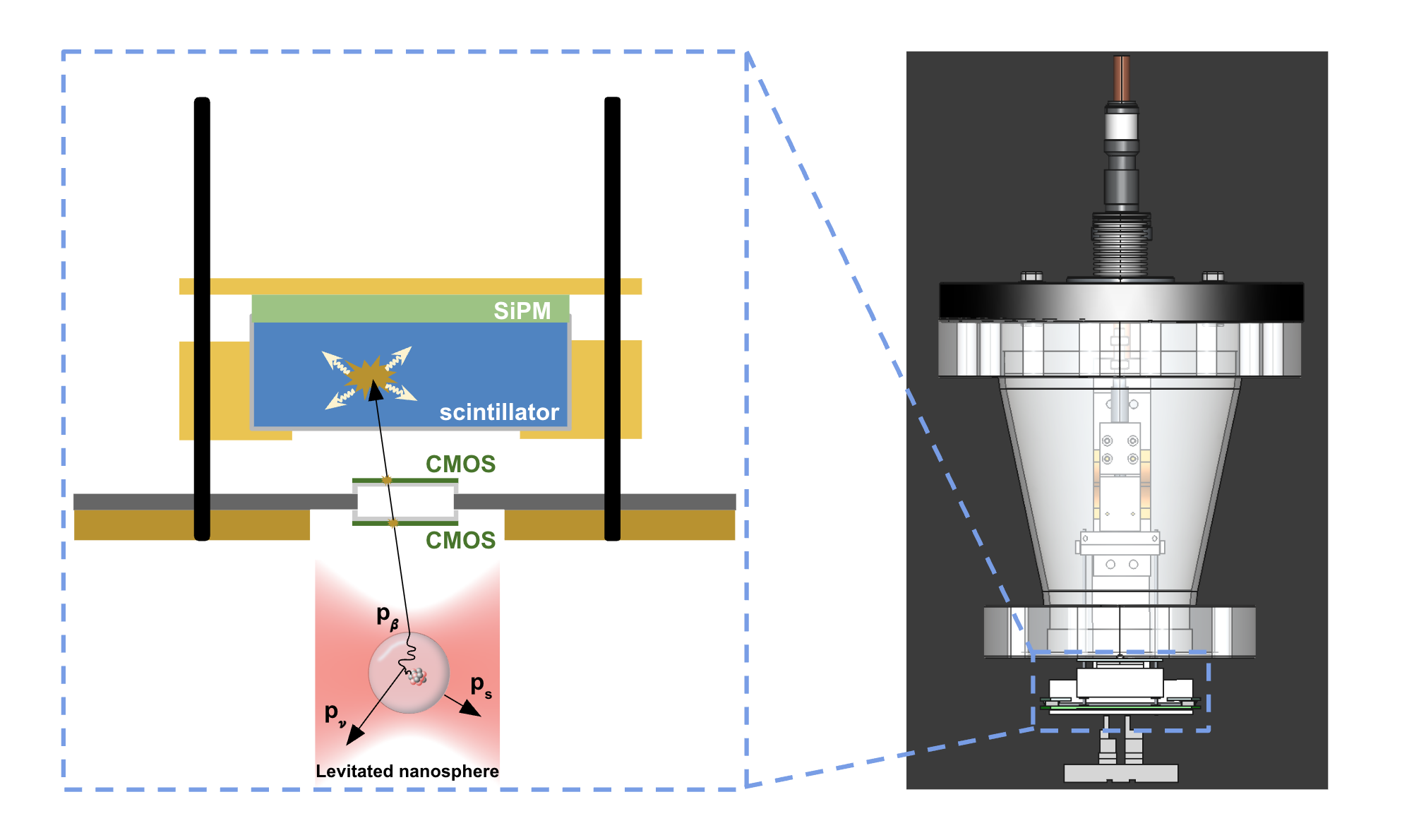}
	\caption{\emph{Left:} Schematic of the QuIPS system (not to scale). A $\beta$-decay within an optically levitated nanosphere emits a $\beta$ particle (momentum $p_\beta$), a neutrino ($p_\nu$), and imparts a recoil impulse to the sphere ($p_s$). The $\beta$ particle traverses two parallel thinned CMOS sensors for directional tracking before entering the scintillator, where it deposits its energy; the resulting scintillation light is read out by the SiPM array. Conservation of momentum on the center-of-mass of the nanosphere allows the neutrino momentum to be inferred from the measured $\beta$ momentum and the sphere recoil. \emph{Right:} CAD rendering of the as-built detector assembly.}
	\label{fig:Schematic}
\end{figure}

The QuIPS electron detector is unique in combining event-by-event tracking and calorimetry of electrons from hundreds of keV to minimally ionizing MeV energies, all within the UHV environment immediately surrounding the optical trap. In the context of optomechanical experiments, the deployment of a particle detector around a levitated nanosphere is, to our knowledge, unprecedented. We meet the requirements outlined in Sec.~\ref{sec:intro} with a detector whose tracking is handled by two planes of parallel CMOS sensors and whose calorimetry is performed by a scintillator backing detector read out by SiPMs. A schematic and CAD rendering are shown in Fig.~\ref{fig:Schematic}, and images of the as-built detectors in Fig.~\ref{fig:detImage}. The following subsections describe the tracking and calorimetric detectors in turn.

\subsection{CMOS Tracking Detectors}
\label{sec:tracking}

The tracking system serves two purposes. First, it provides the directional measurement of the $\beta$ particle emitted from the nanosphere, which is required to reconstruct the full $\beta$ momentum vector. Second, it enables vertexing of the $\beta$ trajectory back to its point of origin, a capability essential for background rejection. Electrons that scatter off the lenses and other material used to form the optical trap, as well as cosmic ray muons, produce tracks that can mimic genuine $\beta$ interactions. By reconstructing the trajectory of an ionizing particle from hits in at least two parallel tracking planes, these backgrounds can be rejected: only particles whose extrapolated trajectory points back to the point-like nanosphere source are retained.

The tracking subsystem consists of two LBNL custom-designed active pixel sensors, fabricated in a 180~nm CMOS image sensor process with a 3T pixel architecture. Each pixel contains three transistors: one for reset, one in a source-follower configuration for voltage measurement, and one to select the pixel and row for readout~\cite{Lee1995AnAP}. These sensors feature a pixel pitch of 4.58~$\mu$m and an array of 896~$\times$~896 pixels, corresponding to a full sensor area of approximately 16.8~mm$^{2}$. The sensors are read out at a frame rate of 250 Hz, which is sufficient for the expected beta-decay rates of order a few Hz from a single nanosphere, depending on the number of loaded atoms and the isotope's half-life; the frame rate can be increased if required for operation with higher-activity sources. Each chip also contains two different types of pixels with slightly different capacitance due to the inclusion of microfabricated guard rings. 

A custom readout board was developed at Lawrence Berkeley National Laboratory (LBNL) to operate both CMOS sensors on a single board, enabling simultaneous, synchronized readout of the two tracking planes with a single set of cables, bias supplies, and digital interfaces. The board provides the necessary bias voltages, clock signals, and digital readout interfaces for both sensors. In this prototype, consolidation is achieved by reading out only the central half of each sensor rather than its full pixel array, giving an instrumented active area of approximately 8.42~mm$^2$ per CMOS. Because the two planes are read out in parallel, this halving keeps the total data volume and the number of readout channels equal to that of a single full sensor, allowing the dual-plane tracker to reuse the readout architecture, cabling, and data-acquisition infrastructure designed for a single chip without duplication. This does, however, adversely impact the geometric acceptance of $\beta$ particles traversing the CMOS. Future iterations of the readout will instrument the entire pixel array of each sensor, recovering the full active area and improving the geometric acceptance. 

\begin{figure}[!t]
    \centering
    \includegraphics[width=0.85\linewidth]{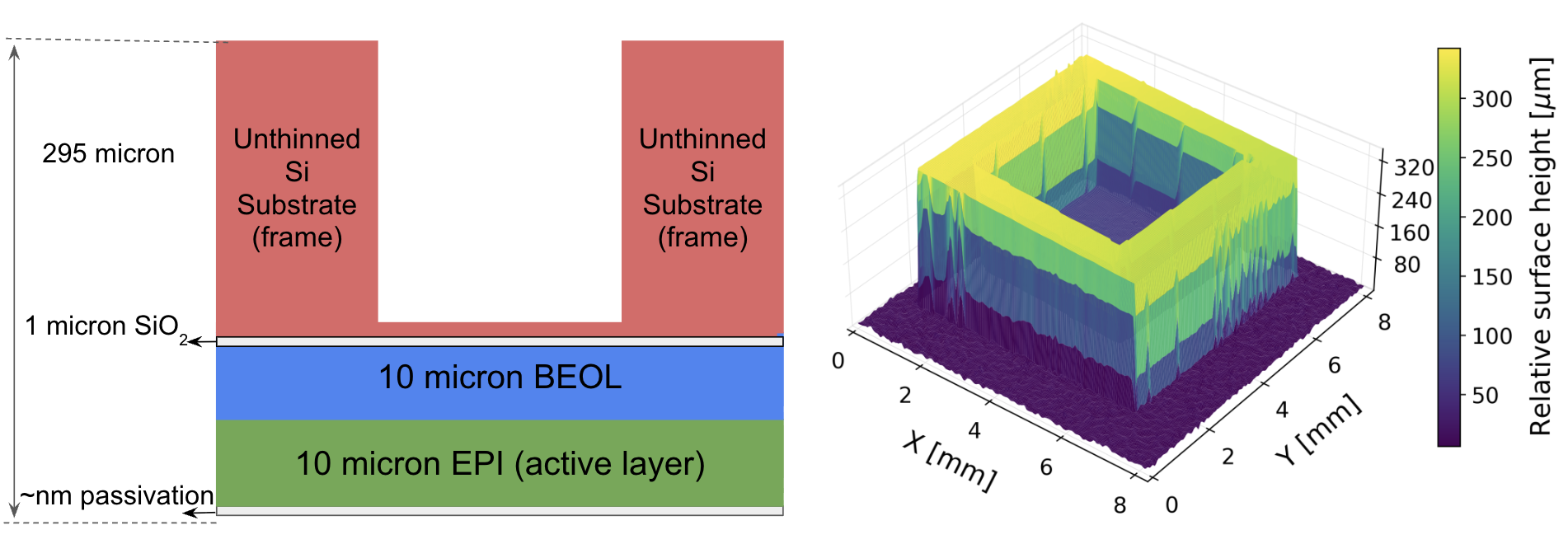}
    \caption{\emph{Left:} Schematic of the cross section of the thinned CMOS sensor. The silicon substrate is removed behind the fabricated region down to the SiO$_2$ layer, leaving the active epitaxial layer, BEOL layer, and SiO$_2$
    passivation layer in the electron path. \emph{Right:} A 3D surface scan of the CMOS sensor, showing the thinned area and the unthinned frame.}
    \label{fig:CMOS_xsec}
\end{figure}

The chips are backside illuminated (BSI) with each chip consisting of, in sequence, a 10~$\mu$m thick active epitaxial layer, a 10~$\mu$m thick back-end-of-line (BEOL) layer for circuitry, a 1~$\mu$m thick SiO$_{2}$ passivation layer, and 275~$\mu$m of silicon substrate. The standard CMOS silicon wafer thickness of 275~$\mu$m would introduce excessive energy loss and multiple scattering for the electrons, significantly degrading vertexing of the electron's path. Thus, the sensors are thinned in-house at LBNL employing a deep reactive-ion etching (DRIE) process. As shown in Fig.~\ref{fig:CMOS_xsec}, the 275~$\mu$m silicon substrate is removed behind the fabricated region down to the SiO$_2$ layer, leaving the 10~$\mu$m active epitaxial layer, 10~$\mu$m BEOL layer, and 1~$\mu$m SiO$_2$ passivation layer as the material traversed by the electron. The silicon surrounding the active area is retained to provide a frame by which the chip is mounted to the readout board.

\subsection{Scintillator-SiPM Calorimeter}
\label{sec:calorimeter}

The calorimeter must reconstruct the energy of electrons emitted following a $\beta$-decay, spanning the range from tens of keV up to the $^{90}$Y endpoint of 2.279~MeV. Since the energy measurement determines the magnitude of the electron momentum, any undetected energy loss biases the inferred neutrino momentum. An underestimate of the electron energy translates directly into an overestimate of the missing momentum carried by the neutrino, mimicking the signature of a heavy sterile neutrino. The primary energy loss mechanisms are two-fold. The first is bremsstrahlung radiation emitted as the electron ranges out in the scintillator. These photons can escape the active volume before depositing their energy. The second process is backscattering, in which an electron undergoes a large-angle scatter near the entrance face and exits the scintillator before depositing its full energy. The cross-sections for both effects scale strongly with the atomic number of the stopping material, setting a requirement that the calorimeter be constructed from a low-Z scintillator, while retaining sufficient light yield for adequate energy resolution. Its active volume must also be large enough to fully stop electrons up to the $^{90}$Y endpoint.

We use an EJ-200 plastic scintillator, a polyvinyltoluene (PVT)-based material composed primarily of hydrogen and carbon, whose low effective atomic number directly suppresses both bremsstrahlung and backscattering losses. EJ-200 is quoted as having a light yield of 10,000~photons/MeV and a peak emission wavelength of 425~nm \cite{Eljen_EJ200}. It has also been shown to have a linear energy response from hundreds of keV to a few MeV~\cite{Tran2018,LSwiderski_2012}, encompassing the bulk of the $^{90}$Y $\beta$ spectrum. The scintillator is 26.0~mm~$\times$~26.0~mm~$\times$~15.0~mm. The 15.0~mm depth along the direction of electron incidence is chosen to ensure that a $^{90}$Y $\beta$ particle at the full $Q$-value endpoint of 2.279~MeV ranges out completely within the active volume. The characteristic stopping range of a 2.3~MeV electron in PVT is approximately 11~mm. The transverse dimensions are chosen to capture all angles of electrons passing through the two CMOS planes while maintaining a compact form factor.

The scintillation light is read out by a 16-channel Hamamatsu S14161-6050HS SiPM array. Its photon detection efficiency peaks at approximately 50\% near 425~nm, closely matched to the peak emission wavelength of EJ-200~\cite{hamamatsu_S14161}, maximizing the detected light yield and thus the energy resolution. Each channel contains 14,331 microcells, enabling a linear energy response out to several MeV without the onset of saturation due to microcell pileup. The physical dimensions of the 16-channel array are well-matched to the 26.0~mm~$\times$~26.0~mm face of the EJ-200 scintillator, ensuring high geometric coverage and uniform light collection over the readout surface. The primary drawback of this device is its elevated dark count rate at room temperature, which raises the effective energy threshold for low-energy $\beta$ detection. Dark count rates can be reduced by several orders of magnitude through modest cooling of the SiPM operating temperature~\cite{NepomukOtte:2016ktf}.

Light collection within the scintillator-SiPM assembly is enhanced by wrapping all faces of the scintillator, except the SiPM-coupled readout plane, with aluminum foil to provide reflectivity. The face directed toward the CMOS tracking sensors—through which the $\beta$ particle enters—uses thinner foil (10~$\mu$m) than the remaining faces ($>$50~$\mu$m), minimizing the dead material the electron must traverse and thereby reducing the energy lost before the active volume. The entire assembly is housed within a light-tight PEEK enclosure, which has an open face directed toward the CMOS tracking sensors, through which $\beta$ particles enter after traversing the two CMOS planes.

\subsection{Mechanical, Vacuum and Thermal System}
\label{sec:thermal}

Compatibility with the levitated optomechanical setup imposes stringent constraints on the detector geometry, UHV compatibility, and thermal performance.

\begin{figure}[!t]
  \raggedright
  % Left Column: Two stacked images
  \begin{minipage}[b]{0.445\textwidth}
    \begin{subfigure}{\linewidth}
      \centering
      \includegraphics[width=.8\linewidth]{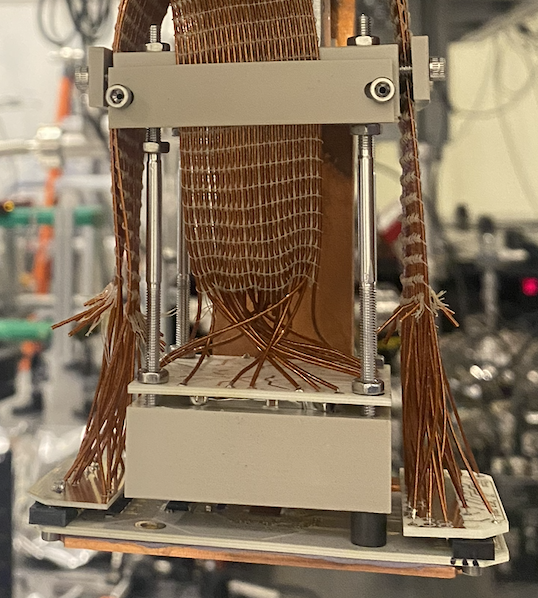}
    \end{subfigure}
    \\[2ex] % Vertical space between them
    \begin{subfigure}{\linewidth}
      \centering
      \includegraphics[width=.8\linewidth]{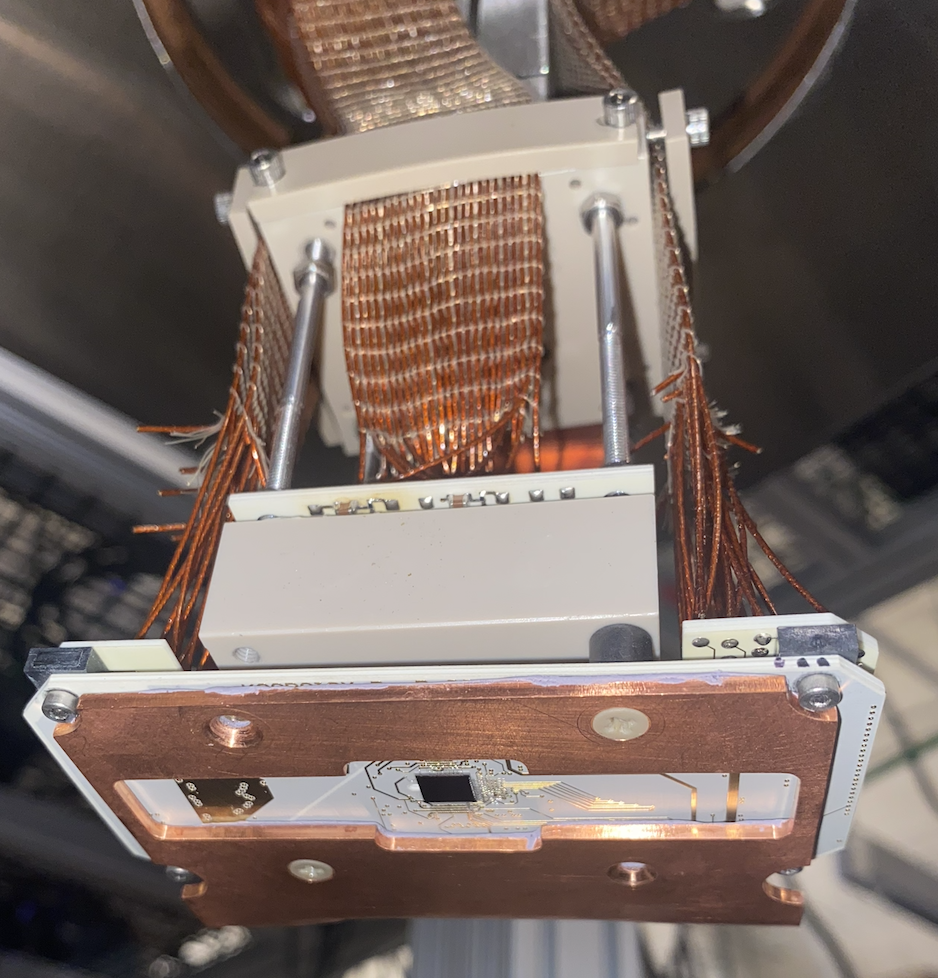}
    \end{subfigure}
  \end{minipage}
  % Right Column: One large image
  \begin{subfigure}[b]{0.5\textwidth}
    \centering
    \includegraphics[width=1.0\linewidth]{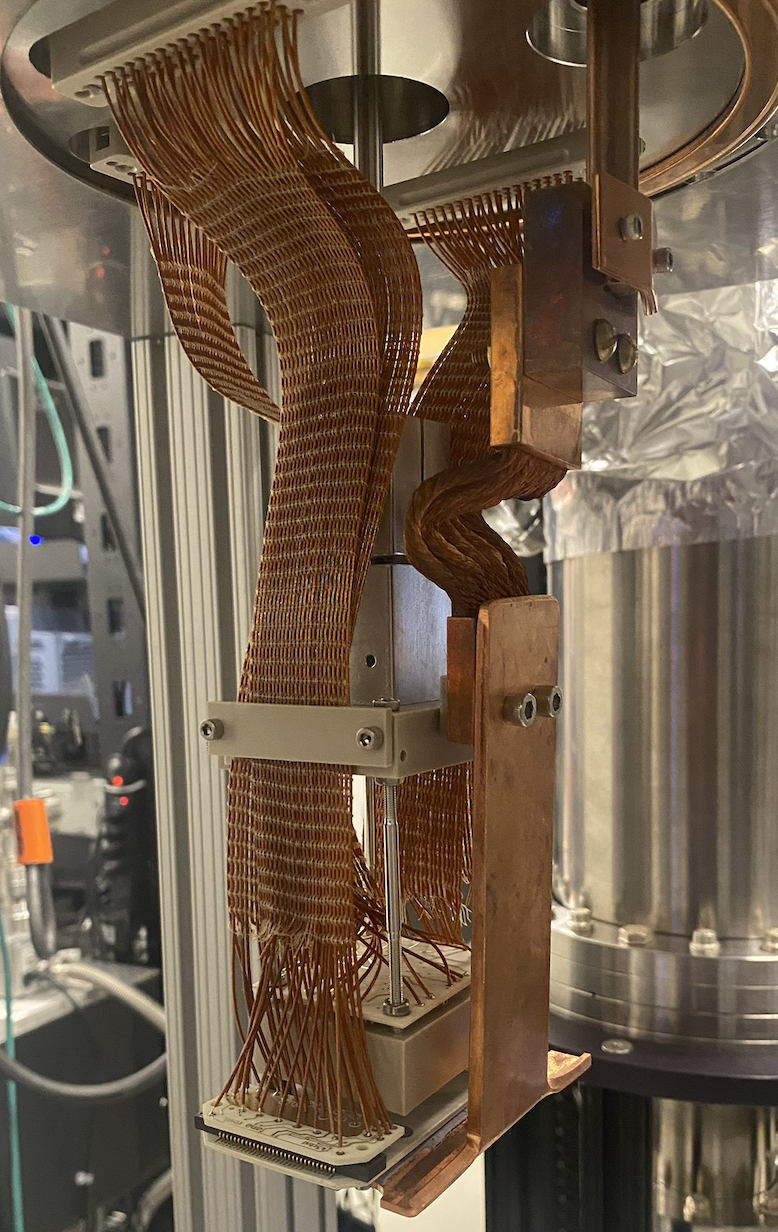}
  \end{subfigure}
  \caption{\emph{Left:} Photographs of the as-built QuIPS detector assembly. The images show the integrated CMOS tracker and scintillator-SiPM calorimeter mounted on their support structure, including the copper thermal-management straps used to conduct heat away from the sensors and SiPMs in vacuum. \emph{Right:} The full assembly installed in the vacuum chamber, hanging from an 8-inch conflat flange.}
  \label{fig:detImage}
\end{figure}

The mechanical design of the detector is shown in Fig.~\ref{fig:Schematic}, with the as-built setup shown in Fig.~\ref{fig:detImage}. The CMOS board is placed above the optical lenses, with the CMOS sensors centered on the nanosphere position. The distance between the nanosphere and the CMOS sensors is minimized to maximize the solid-angle coverage for emitted $\beta$ particles. The PEEK scintillator holder, with the scintillator housed inside, is placed above the CMOS board, giving a separation of 6.5~mm between the scintillator-side CMOS sensor and the scintillator to allow space for the surface-mount components on the CMOS board. The SiPM array, followed by its readout board, is pressed against the scintillator for efficient light collection. The full assembly is aligned and held by a bellows-mounted platform, which provides two inches of vertical travel for integration of the $\beta$-particle detector with the optomechanical setup.

During operation in vacuum, the CMOS sensors, SiPMs, and inline circuit components dissipate heat that must be removed. Otherwise, the sensor temperature can rise uncontrollably, increasing noise and potentially rendering the sensors inoperable. In the absence of convective cooling, heat can only be removed by conduction through the mechanical supports and by radiation. We address this with a dedicated conductive cooling path from atmosphere into the vacuum chamber, formed by a 1.33-inch diameter copper rod, as shown in the CAD rendering in Fig.~\ref{fig:Schematic}. The copper rod connects to a flexible copper strap, which in turn connects to a rigid 0.08-inch-thick copper L-bracket, as depicted in Fig.~\ref{fig:detImage}. The L-bracket is attached to the underside of the CMOS board with a thermally conductive, vacuum-compatible adhesive, providing both mechanical support and passive cooling for the CMOS.  A
16.0~mm~$\times$~16.0~mm window is opened at the center of the L-bracket to clear the path of the electrons. The thermal performance is characterized under normal operating conditions using the temperature sensor integrated on the CMOS board near the CMOS sensors. With this passive connection alone, the CMOS temperature remains below 35$^{\circ}$C in vacuum, within the operational range and with limited impact on the thermal noise. An air-side cooling system is deployed to provide additional cooling power. The air-side end of the 1.33-inch-diameter copper rod is clamped in a copper block, which is coupled through a thermoelectric cooler (TEC) to a liquid-cooled heat sink. At approximately one-third of the TEC's rated maximum power, the copper block is cooled below 0$^{\circ}$C and the CMOS temperature is correspondingly kept below 17$^{\circ}$C.

The entire detector must operate within ultra-high vacuum ($<10^{-7}$~mbar), a requirement imposed by the levitated optomechanical setup. The materials used throughout the detector subsystems are therefore selected for vacuum compatibility. Low-outgassing materials are used wherever possible: PEEK for structural components, such as the scintillator holder and mechanical platform; Kapton for cabling; and low-outgassing Rogers laminates for the boards on which the CMOS sensors and SiPMs are mounted. Copper and stainless steel are used for the remaining parts. With this material selection, a base pressure of $6\times10^{-8}$~mbar was achieved. We attribute the limiting factors to residual outgassing and low-conductance regions associated with the screw geometry and the plastic-packaged surface-mount components on the CMOS board.

\section{Calibration}
\label{sec:calibration}

This section details the operation and calibration of the tracking and calorimeter subsystems individually, establishing the basis for their combined operation.

\subsection{CMOS Tracking Detectors}
\label{sec:cmos_cal}

The CMOS sensors read out 896~$\times$~896 pixels through 16 parallel channels (i.e. 56 columns of pixels), served by 4 ADC chips with 4 dedicated channels each. The frame rate is 222~Hz corresponding to a 4~ms exposure and 0.5~ms frame readout: during which the sensors are still acquiring exposure. Two different readout modes are available. In one mode a global reset is applied to all pixels after every frame, resetting the accumulated charge before the next exposure begins. In the other, the global reset is suppressed for a programmable number of consecutive frames, typically 100, before the reset is issued. This suppression of the per-frame reset enables correlated double sampling (CDS), in which the signal in a given frame is computed as the difference between the current frame and the immediately preceding frame. CDS eliminates the per-pixel charge noise (or kTC noise) from the Johnson noise on a pixel's storage capacitor that is otherwise resampled after each reset and cannot be subtracted without a correlated reference. The suppression of this noise via CDS is an essential feature for achieving sensitivity to the few-keV energy depositions expected from minimally ionizing particles traversing the 10~$\mu$m active epitaxial layer. 

\subsubsection{Noise Profiling with Dark Data}
\label{sec:noise}

\begin{figure}[!t]
	\centering
    \includegraphics[width=0.6\columnwidth]{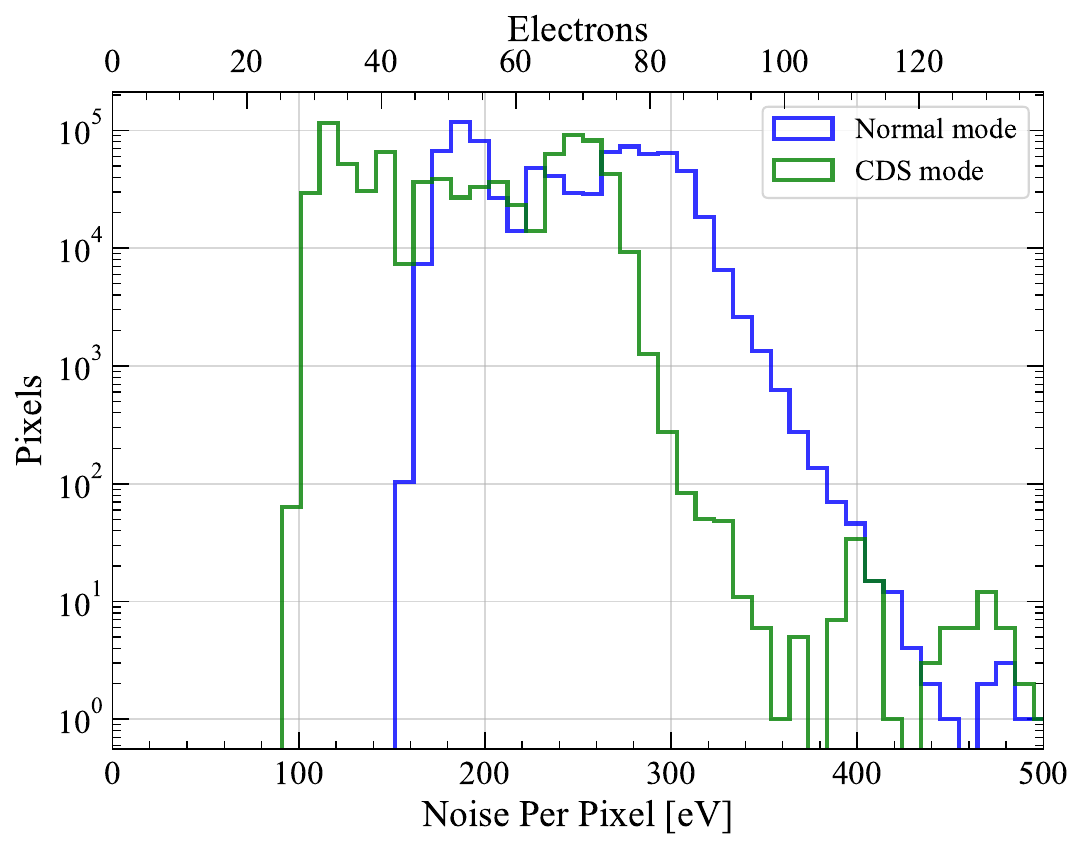}
	\caption{The per-pixel standard deviation is shown for dark data taken with (green) and without (blue) correlated double sampling applied. The per-pixel response is converted into energy units using the calibration discussed in this work.}
	\label{fig:Noise}
\end{figure}

The noise performance of the CMOS sensors was characterized using dark data taken with no radioactive source present and no external trigger, allowing the intrinsic noise floor to be measured independently of any particle signal. The processing pipeline proceeds as follows. First, CDS is applied by subtracting consecutive frames within each reset period. Frames coinciding with global reset events are identified and discarded. For the comparison with dark data acquired in normal mode, this first step is omitted. Next, the DC baseline offset of each pixel, its median response across all frames, is subtracted to remove pixel-to-pixel pedestal variations. Finally, a correlated noise removal step is applied. This step targets a low-frequency tone common to all channels within a single readout row present due to the simultaneous parallel readout of the 16 channels. This correlated readout noise manifests as horizontal striations in the two-dimensional pixel map and is estimated on an event-by-event basis by computing the per-row median signal across all pixels within a channel and subtracting it.

The noise profile is computed by taking the standard deviation of each pixel across more than 1000 frames in a dark environment. Following the procedure discussed above, Fig.~\ref{fig:Noise} shows the per-pixel noise with CDS applied (green) and without CDS applied (blue) under the same conditions. The spectrum is converted into energy units using the $^{55}$Fe-derived per-channel gain calibration discussed in the next section. Without the application of the decorrelation step, the noise would be approximately a factor of two larger than shown. The application of CDS reduces median overall noise by 22\%, corresponding to 14~electrons or 49~eV. The noise floor corresponds to approximately 33--70~electrons, equivalently 120--254~eV, per pixel. The remaining noise contributions are primarily from thermal noise as these CMOS are operated near room temperature.

The noise profile was also compared between dark data and data taken with a 53~nCi $^{90}$Sr/$^{90}$Y button source and an approximately GBq $^{55}$Fe source. In the case of the former, the measured noise was consistent with dark data within statistical uncertainties, whereas the latter case was subject to increased noise due to the high rate source. The noise-threshold analysis is therefore robust unless the source rate is too high.

To identify genuine particle interactions, as opposed to upward fluctuations in the noise distribution, a detection threshold is set as a multiple of the per-pixel noise in terms of $\sigma$. Given that the per-pixel noise is well approximated as Gaussian and that there are 401408 pixels per CMOS sensor, the probability that at least one pixel in a given frame fluctuates above a 5$\sigma$(6$\sigma$) threshold is 11\%($4\cdot10^{-5}~\%$). Thus, raising the threshold reduces the false positive rate at the cost of a slightly higher energy threshold. For the primary physics data-taking runs described in Sec.~\ref{sec:cal}, a 6$\sigma$ threshold is adopted as a compromise between threshold and false positive rejection. Physical events can consist of either a single pixel above threshold or a cluster of neighboring pixels above threshold.

\subsubsection{Gain Calibration with $^{55}$Fe}
\label{sec:Fe55}

\begin{figure}[!t]
	\centering
    \includegraphics[width=0.6\columnwidth]{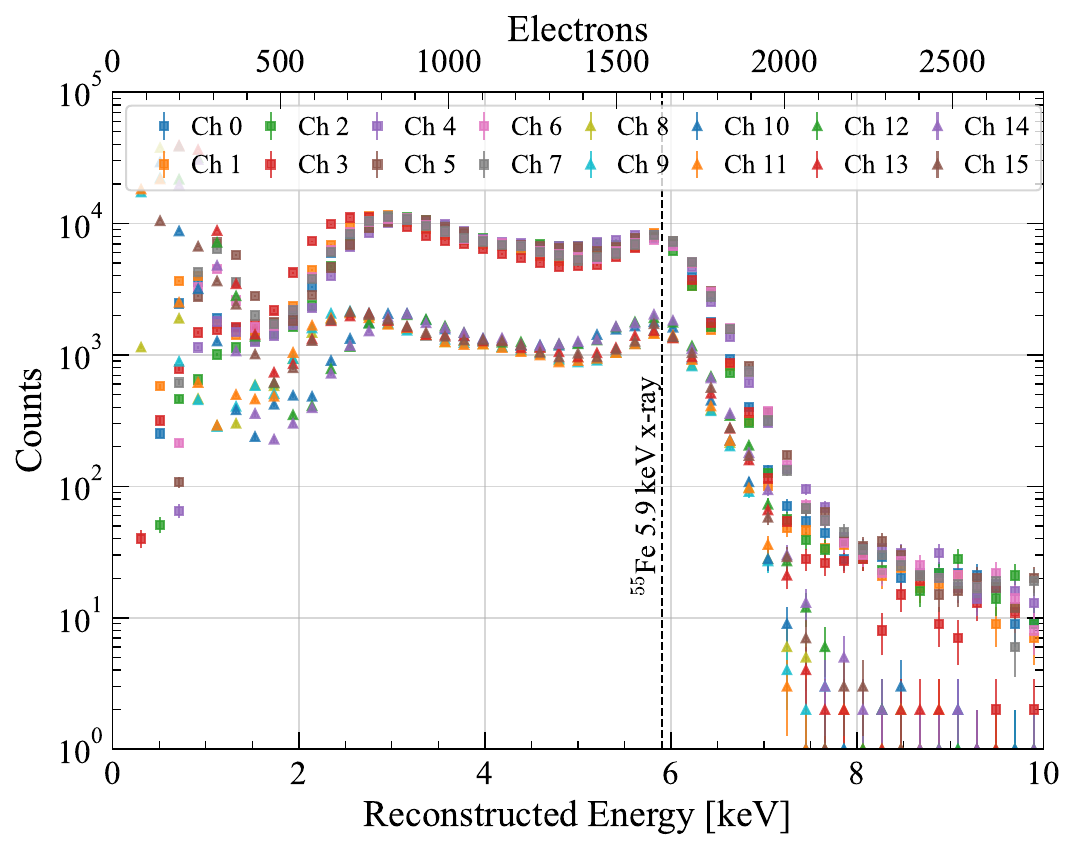}
	\caption{The per-channel calibrated response to $^{55}$Fe 5.9~keV x-rays. The source is placed facing the trap-side CMOS (channels 0-7). Thus, there are fewer x-rays that reach the scintillator-side CMOS (channels 8-15).}
	\label{fig:Fe55}
\end{figure}

The energy scale of the CMOS sensors is established using approximately GBq $^{55}$Fe source, which produces 5.9~keV X-rays via electron capture to excited states of $^{55}$Mn. Each 5.9~keV X-ray absorbed in silicon produces approximately $N_{e^-} = E/W = 1634$ electron-hole pairs, where the mean energy required to produce one electron-hole pair in silicon at room temperature is $W = 3.61$~eV~\cite{PhysRevLett.56.2195}. Data were collected with the $^{55}$Fe source positioned facing the trap-side CMOS sensor. The mean free path of the 5.9~keV X-ray in silicon is approximately $28~\mu$m, which is long enough to provide sufficient statistics to also calibrate the scintillator-side CMOS in this configuration. This is readily confirmed in the data.

Channel-to-channel variation in the gain is observed even among channels of the same nominal gain mode, arising from transistor-to-transistor variation in the source follower circuits that amplify the pixel signal before digitization. This variation makes it essential to calibrate each of the 16 readout channels individually. The $^{55}$Fe peak position in ADCC counts is identified for each channel by fitting the photopeak with a Gaussian plus linear background. Figure~\ref{fig:Fe55} displays the calibrated response across the sixteen channels using single-pixel hits above a 3$\sigma$ per-pixel threshold. The $^{55}$Fe spectra from all 16 channels align at 1634 electrons (5.9~keV), confirming the calibration procedure. The high-gain channels are a factor of approximately 1.6 more sensitive per ADCC than the low-gain channels, consistent with the sensor design. This calibration result is independent of the readout mode of the CMOS which only impacts the measured per-pixel noise.

The energy calibration allows us to convert the noise profile into meaningful units as shown in Fig.~\ref{fig:Noise}. Minimally ionizing particles will deposit, on average, 3--4~keV in 10~$\mu$m of silicon. Given the noise is $<250$~eV in 86\% of the pixels, a conservatively high threshold of 6$\sigma$ can be employed without significant loss of events, or acceptance of false positives.

\subsection{Scintillator-SiPM Calorimeter}
\label{sec:scint_cal}

Scintillation pulses produced by energy depositions in the scintillator are collected by a 16 channel Hamamatsu S14161-6050HS SiPM. The active surface of the SiPM is pressed against the scintillator without any optical grease due to concerns of incompatibility with high vacuum. The SiPM is read out by an 8~channel CAEN DT5730 digitizer with a sample rate of 500~MS/s. The 16 channels of the SiPM are grouped into 8 channels to match the digitizer inputs: the inner 4 channels are read out individually, and the outer 12 channels are ganged, in parallel, into groups of 3, centered around the corners. The signal is read out directly on the CAEN digitizer, terminated on a 50~$\Omega$ line. The bias line includes two low-pass filters, one on the SiPM board in vacuum and the other on an air-side breakout box. The total inline resistance is 1.55~M$\Omega$, and each capacitor is 5~$\mu$F. The remainder of this section details the calibration of the SiPM and the scintillator-SiPM system. 

\subsubsection{Energy Calibration}
\label{sec:scint_Ecal}

\begin{figure}[!t]
    \centering
    \includegraphics[width=\linewidth]{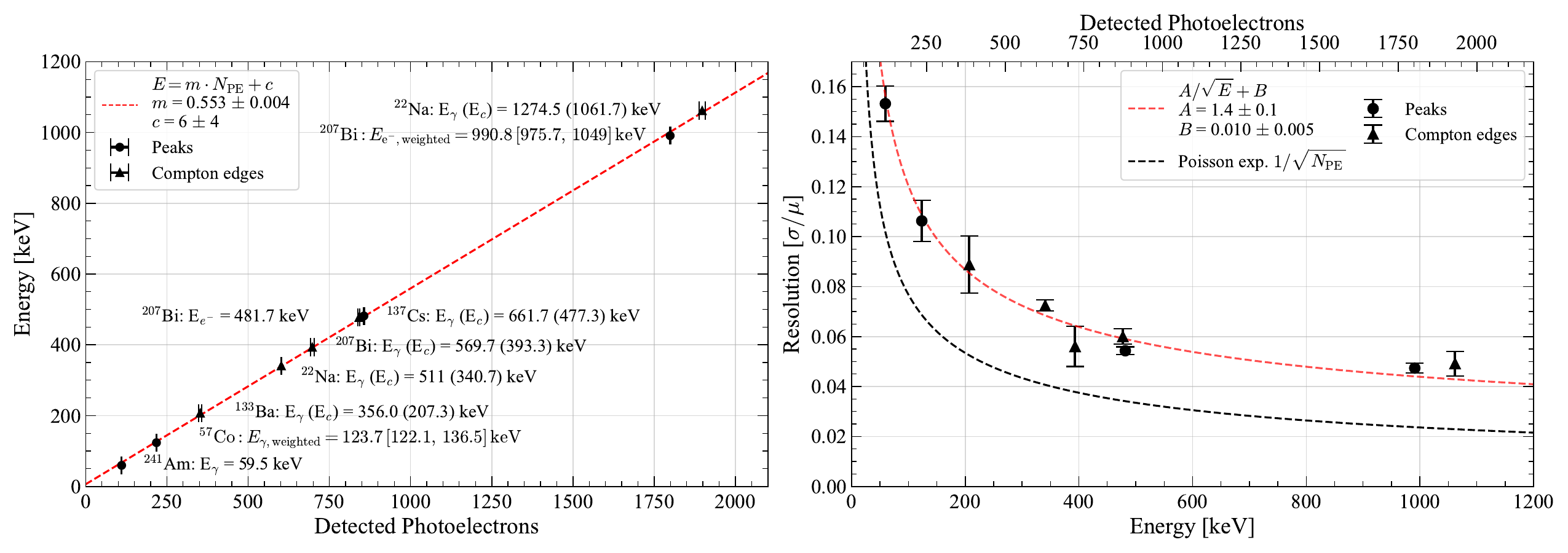}
    \caption{\emph{Left:} Energy calibration of the EJ-200 scintillator-SiPM detector using photopeaks, Compton edges, and $^{207}$Bi conversion electron features spanning energies of 60~keV to 1062~keV. The calibration is well-matched to a linear response. \emph{Right:} Measured fractional energy resolution of the detector compared to the limiting Poisson value $1/\sqrt{N_{\mathrm{PE}}}$. A measured energy resolution of 5\% is measured using the $\sim1$~MeV conversion electrons from $^{207}$Bi.}
    \label{fig:ScintCal}
\end{figure}

The response of the Hamamatsu S14161-6050HS SiPM array to single photoelectrons (PE) was characterized. The SiPM has a manufacturer-quoted breakdown voltage of approximately 38~V. The bias applied to the circuit was adjusted to account for the voltage drop across the 1.55~M$\Omega$ total resistance. The bias applied to the SiPMs throughout all datasets was 43.4~V, corresponding to an overvoltage of 3.1~V. The single PE size in the individual channels was measured using random event windows and confirmed to have a gain consistent with the manufacturer specification~\cite{hamamatsu_S14161}. The ganged channels were biased at the same overvoltage, such that the individual SiPMs experience the same gain. Though the single PE area is invariant, the increased total capacitance led to an increased fall time in the single PE pulses. This analysis relies on single PE heights since this quantity is insensitive to the relatively high afterpulsing rate of these SiPMs. Thus, the single PE height of both individual and ganged channels was directly measured and confirmed to be related by the change in the pulse shape. This analysis provides a factor to convert from pulse height to PE per digitizer channel. Using the manufacturer specification~\cite{hamamatsu_S14161}, the gain in each dataset was used to determine the photon detection efficiency and crosstalk probability, and was subsequently normalized to a 3.1~V overvoltage. 

The energy calibration of the scintillator–SiPM system was performed using a suite of radioactive sources spanning energies from 60~keV to 1062~keV. Because the EJ-200 plastic scintillator is a low-$Z$ medium in which Compton scattering dominates over photoabsorption above approximately 100~keV, the calibration spectra are dominated by Compton features rather than full-energy photopeaks. The location and uncertainty of the Compton edges are estimated using the method of Ref.~\cite{Safari:2016ypc}, which models these features as a second-order polynomial convolved with a Gaussian detector resolution.

To correct for position-dependent light collection in the scintillator, a weighted centroid location is calculated for each event using the number of photoelectrons measured in each channel. A correction factor is then applied based on a lookup table constructed from a $5 \times 5$ binning of centroid positions. The lookup table was generated using $^{207}$Bi 1~MeV conversion-electrons by measuring the percentage shift of the peak position relative to the central bin.

Small nonlinearities in EJ-200 light yield are expected at the lowest energies probed due to ionization quenching well described by Birks' law~\cite{CRAUN1970239}. Quenching corrections based on Ref.~\cite{TAJUDIN2020109086} are introduced to maintain energy linearity in this calibration. Corrections are as large as 15\% at 60~keV. Although the highest available calibration data point is the 1062~keV Compton edge from $^{22}$Na, the light yield of EJ-200 in the 1--2.3~MeV energy range relevant to $^{90}$Y is well established to be nearly constant~\cite{Nassalski_2008,Payne_2011,Tran2018,LSwiderski_2012}. Thus, extrapolating this linearity in the regime relevant to $^{90}$Y from 1062 keV to 2279~keV is justified. Further, the high microcell count (14331/channel) ensures that any non-linearity in the energy response due to scintillation photon pileup in the SiPM is constrained to be less than a few percent at 2.3~MeV.

The energy calibration data and resulting fit are shown in left plot of Fig.~\ref{fig:ScintCal}. The resulting calibration points are well described by a linear relationship between the number of detected photoelectrons $N_\text{PE}$ and the deposited energy $E~[\text{keV}] = m\cdot N_\text{PE} + c$ with $m = 0.553 \pm 0.004$~keV/PE\ and $c = 6 \pm 4$~keV from a fit to all calibration points. 

The measured energy resolution, $\sigma/\mu$, is shown in the right plot Fig.~\ref{fig:ScintCal} comparing the measured detector resolution to the Poisson expectation based purely on the mean number of detected photoelectrons. The excess resolution beyond the photon-counting limit is attributed to position-dependent light collection across the scintillator volume either not captured by the coarse corrections applied or dependent on depth in the scintillator. The measured energy resolution at 1~MeV from mono-energetic $^{207}$Bi conversion electrons is 5\%.

We can additionally calculate the geometric light collection efficiency given as the ratio of photons produced to those detected in the SiPMs. The light yield of the EJ-200 is 10,000~photons/MeV~\cite{Eljen_EJ200}. We correct for the 52\% average photon detection efficiency of the SiPMs and 9\% crosstalk probability at our operating voltage to get the number of photons incident on the SiPMs~\cite{hamamatsu_S14161}. We measure an average geometric light collection of $31 \pm 1 \%$.

%%%%%%%%%%%%%%%%%%%%%%%%%%%%%%%%%%%%%%%%
\section{Full System Calibration}
\label{sec:cal}

\begin{figure}[!t]
	\centering
    \includegraphics[width=0.8\columnwidth]{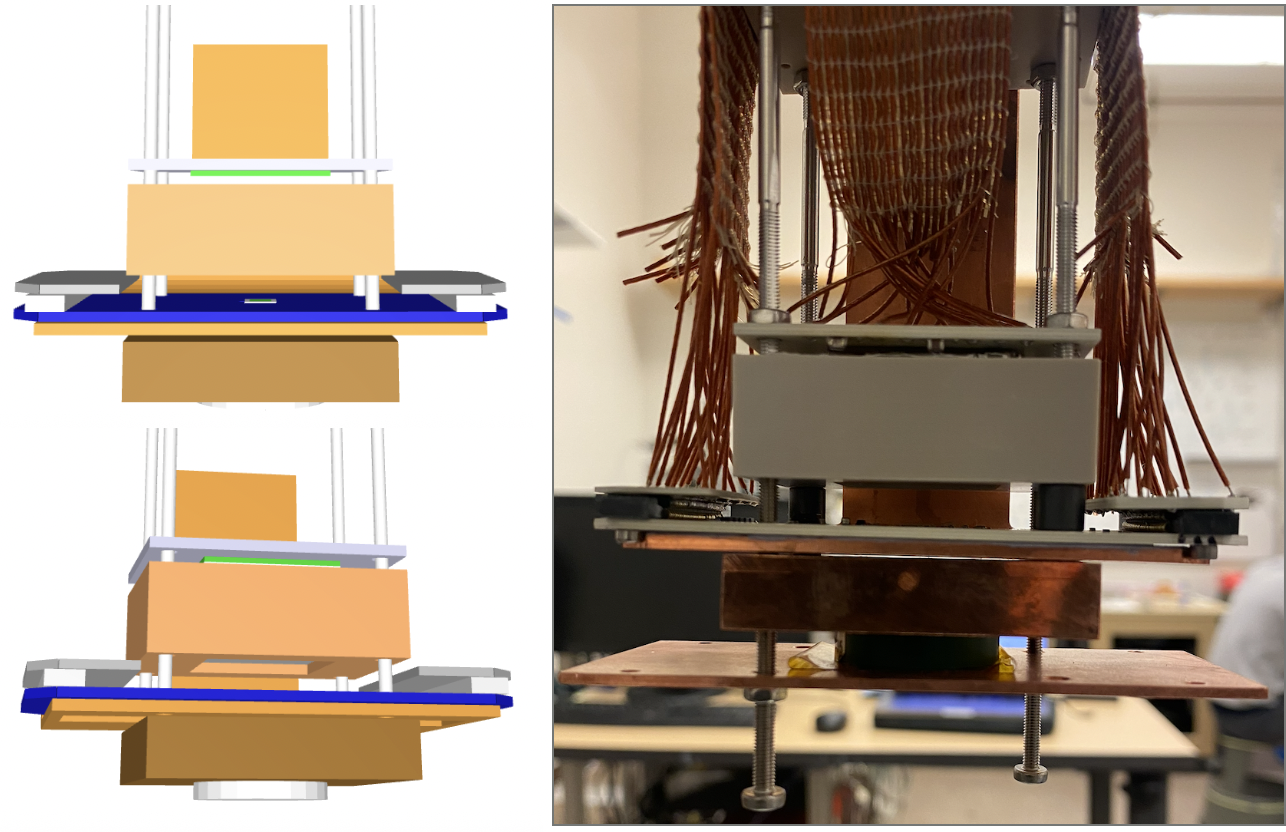}
	\caption{\emph{Left:} A \textsc{Geant4} model of the detector setup for the $^{90}$Sr/$^{90}$Y calibration. \emph{Right:} Image of the detector setup for calibration. A button source is placed facing upward towards the QuIPS electron detector. A 10~mm thick copper block with a 1.5~mm diameter aperture provides collimation for the source incident on the CMOS sensors. }
	\label{fig:calScheme}
\end{figure}

Full system calibration of the QuIPS detector consisted of operating the two CMOS tracking planes and the scintillator–SiPM calorimeter in coincidence with a dedicated operational mode. The system is triggered by events in the scintillator-SiPM system, which in turn produce a gated square pulse sent into the CMOS FPGA. Though the CMOS cycles through exposure and readout continuously, frames are only saved when an external trigger signal from the scintillator-SiPM system is present. This keeps the amount of data stored at manageable levels during high rate and long duration datasets. The FPGA further stores how many triggers are present during an exposure, and at what point during the CMOS frame the trigger occurred (either during exposure or readout). If a trigger occurs during readout, the preceding and/or following frames are stitched together around the row at which the trigger occurred. Timing synchronization between the CMOS and scintillator readout systems is achieved by driving the CAEN DT5730 digitizer with the same 50~MHz oscillator used by the CMOS FPGA.

The data-taking geometry is shown in the right panel Fig.~\ref{fig:calScheme}. A radioactive source is placed beneath a 10~mm thick copper block with a 1.5~mm diameter aperture. This provides a collimated source of electrons towards the trap-side CMOS sensor. The collimator is directly below the copper L-bracket beneath the CMOS board such that the collimator aperture is 1.705~mm from the epitaxial layer of the trap-side CMOS. The left panel of Fig.~\ref{fig:calScheme} shows a custom-built model of the calibration scheme in \textsc{Geant4}. The dimensions in the simulation match the as-built schematics. The \textit{G4EmStandardPhysics\_option4} physics model is employed. In the thin (10~$\mu$m) silicon films, the photoabsorption ionization (PAI) model is used to capture shell corrections for the low energy scattering effects~\cite{WANG20181}. Energy depositions in the active silicon and scintillator volumes have their response smeared using the mean electronic noise shown in Fig.~\ref{fig:Noise} and resolution model in Fig.~\ref{fig:ScintCal}, respectively.

Two separate datasets were acquired: two days of data were with a collimated 53~nCi $^{90}$Sr source and one day without any source present. The latter case being where the scintillator triggers are predominantly from cosmic rays. The scintillator trigger threshold was set at the same value for each dataset. The trigger rate in the scintillator was 0.65~Hz with the source present and 0.51~Hz without the source present, such that the contribution from the $^{90}$Sr/$^{90}$Y electrons is 0.14~Hz.

The analysis of the datasets proceeds by identifying triple coincidences: events in which an above threshold (6$\sigma$) pixel, or cluster of pixels, appears in each of the trap-side and scintillator-side CMOS that is coincident within $\Delta T<1~\mu$s of a triggered scintillator pulse. In cases where there are multiple, spatially distinct clusters in a single CMOS, an additional selection is applied to determine the combination of clusters in each CMOS that best vertex to the location of the emitted electrons, i.e. the collimator.

%% Check math for expectation of muons
A total of 8746 triple coincidences were identified in the 48 hour $^{90}$Sr/$^{90}$Y dataset. The cosmic ray dataset yielded 42 triple coincidence events in just 24 hours. Given an atmospheric muon flux of 70--100 muons/m$^{2}$/s/sr and the CMOS active area of 8.42~mm$^2$, we predict 32--45 muons passing through both sensors per day. This is in agreement with our observation and validates the sensitivity of the CMOS and our algorithmic selection to detect minimally ionizing particles.

\subsection{Data-Driven Detection Efficiency}

%% Data driven efficiency vs energy plot
\begin{figure}[!t]
	\centering
    \includegraphics[width=0.8\linewidth]{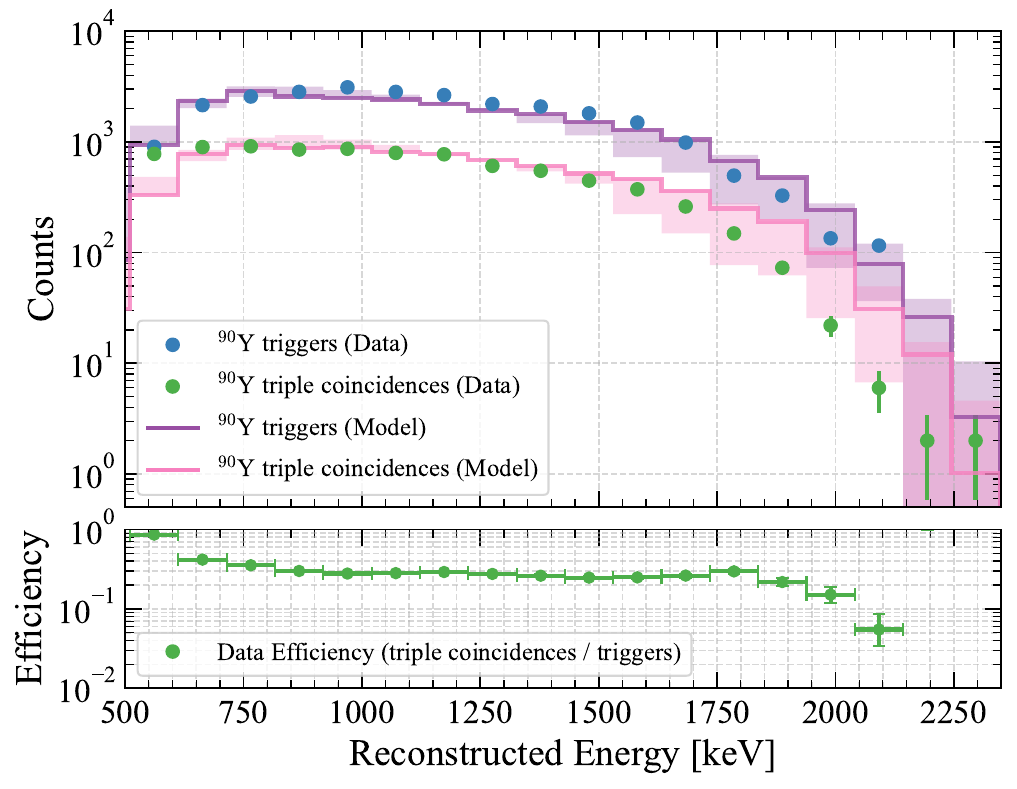}
	\caption{The $^{90}$Sr/$^{90}$Y induced triggers and reconstructed triple coincidences are shown for data (blue and green points) and the \textsc{Geant4} model (purple and pink histograms). The shaded bands represent the uncertainty associated with the detected spectrum based on the source alignment with the collimator (see text for details). The bottom panel shows the ratio of reconstructed triple coincidences as a function of energy. The efficiency is 35\% integrated over the energy range shown. Losses of efficiency are described in the text.}
	\label{fig:eff}
\end{figure}

The detection efficiency of the combined system as a function of $\beta$ energy is extracted in a fully data-driven manner from the $^{90}$Sr/$^{90}$Y and cosmic ray datasets. The denominator for the efficiency is the spectrum of $^{90}$Sr/$^{90}$Y-induced scintillator triggers, obtained by subtracting the no-source (cosmic ray) triggered spectrum from the source triggered spectrum, after normalizing for the respective live times. The numerator is the spectrum of $^{90}$Sr/$^{90}$Y triple coincidences, obtained by subtracting the cosmic ray triple-coincidence spectrum from the source triple-coincidence spectrum. This subtraction removes the contribution of cosmic rays from both quantities, isolating the response of the detector to $\beta$ particles. The efficiency is then the ratio of these two cosmic-subtracted spectra as a function of the reconstructed scintillator energy. 

The top panel of Fig.~\ref{fig:eff} shows the reconstructed trigger (blue points) and triple coincidence (green points) spectra  as compared to their corresponding \textsc{Geant4} model predictions in purple and pink, respectively. The shaded bands represent the uncertainty associated with the source alignment. The bounding cases are created by $\beta$ particles that were produced either within or outside the radial extent of the collimator. In general, the data and model agree well across the energies considered. The lowest energy shown, 500~keV, is set by the conservatively high trigger threshold applied. It is not a fundamental limitation of the system. Given the noise levels observed in the combined system, we believe a trigger threshold down to a few hundred keV is achievable.  

The resulting efficiency is shown in the bottom panel of Fig.~\ref{fig:eff}. The efficiency is approximately flat from 500~keV to 2000~keV. The total efficiency is 35\% above 500~keV. This flatness in efficiency is expected from the underlying detector physics. The stopping power of electrons in silicon varies by less than 10\% between 500~keV and 2000~keV, such that the energy deposited in the 10~$\mu$m active epitaxial layer, and thus the efficiency to reconstruct such tracks, should be approximately constant across this range. We compare the measured efficiency to the dedicated \textsc{Geant4} model. The measured detector resolutions and efficiencies are applied to simulated scintillator and CMOS response. Above 500~keV the simulations predicts an efficiency of 58\%.

The overall scale and energy dependence of the efficiency can be understood as the product of two distinct effects: geometric acceptance and the CMOS energy-threshold. The simulation indicates that, of the events that trigger the scintillator in this energy range, approximately 71\% also geometrically intersect the active region of both CMOS planes. The remaining ~29\% are lost becase those to electrons do not pass through the active area in the scintillator-side CMOS. This effect is due to Coulomb scattering in the silicon and the angle of the electrons emitted from the collimator.

The remaining reduction from this geometric ceiling is driven by the energy threshold applied to the CMOS. Because the detection threshold is defined as a fixed multiple (6$\sigma$) of the per-pixel noise rather than as a fixed energy, the corresponding effective energy threshold differs between the high-gain and low-gain channels and varies across the sensor. An electron must deposit charge above this threshold in both the trap-side and scintillator-side CMOS to be reconstructed. Since the most probable energy deposit in 10~$\mu$m of silicon is only a few keV, a non-negligible fraction of genuine traversals falls below threshold, particularly in the low-gain channels where the effective energy threshold is highest. This threshold effect accounts for the residual gap between the geometric acceptance and the measured efficiency, and it is also the dominant source of the discrepancy between the data and the \textsc{Geant4} prediction, since the simulation does not model charge sharing or charge diffusion across pixels and therefore reproduces the threshold behavior only approximately. 

\begin{figure}[!t]
	\centering
    \includegraphics[width=0.8\linewidth]{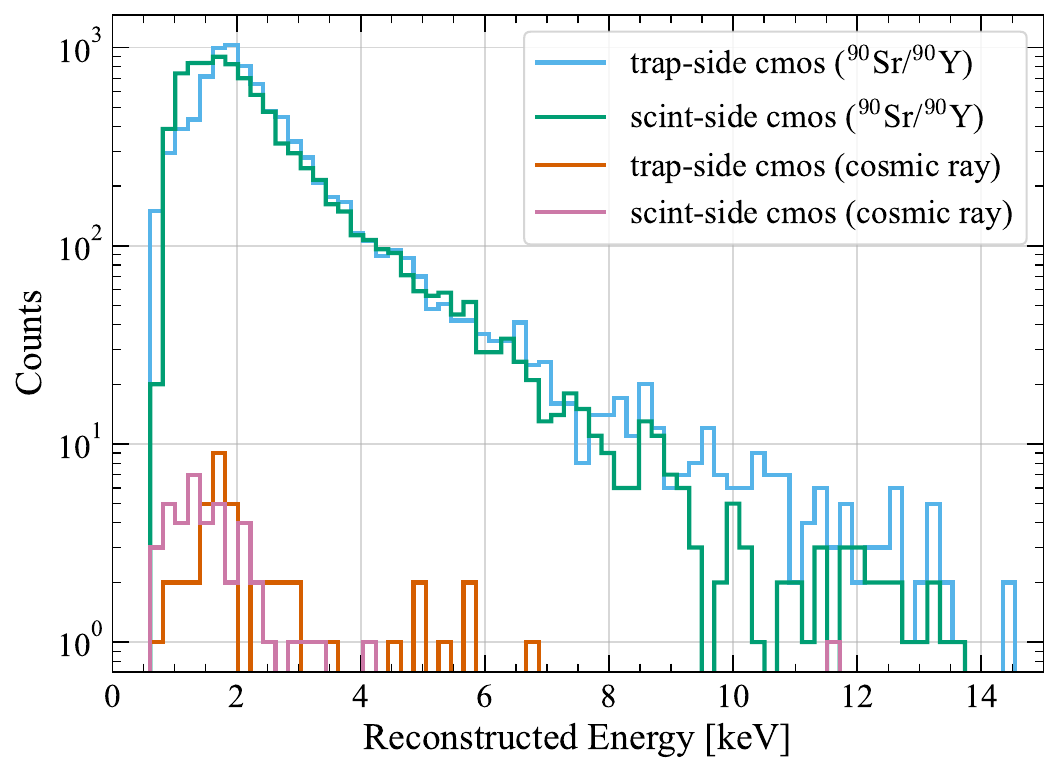}
	\caption{The reconstructed energy deposited in the CMOS sensors by clusters participating in triple coincidences, shown for the trap-side (blue) and scintillator-side (green) planes in the $^{90}$Sr/$^{90}$Y dataset and for the trap-side (orange) and scintillator-side (pink) planes in the cosmic ray dataset. Energies are derived from the $^{55}$Fe per-channel gain calibration.}
	\label{fig:cmos_spectra}
\end{figure}

The energy deposited by the electrons and cosmic rays that produce triple coincidences is shown in Fig.~\ref{fig:cmos_spectra}, separately for the trap-side and scintillator-side CMOS and for both the $^{90}$Sr/$^{90}$Y and cosmic ray datasets. In all cases the reconstructed energy is derived from the $^{55}$Fe-based per-channel gain calibration of Sec.~\ref{sec:Fe55}. The $^{90}$Sr/$^{90}$Y spectra on the two CMOS planes are nearly identical and peak near 2~keV, consistent with the most-probable energy loss expected for a minimally ionizing particle traversing the 10~$\mu$m active epitaxial layer, and exhibit the characteristic Landau tail extending to higher deposited energies. The close agreement between the trap-side and scintillator-side distributions confirms that the two planes respond comparably to the same population of traversing electrons. The cosmic ray triple coincidences, obtained from the no source dataset, populate the same low-energy region as expected of minimally ionizing particles.

This spectrum makes explicit the origin of the CMOS energy-threshold loss discussed above. Because the 6$\sigma$ detection threshold corresponds to an effective energy cut of approximately 1.2~keV, a non-negligible fraction of genuine traversals fall near or below threshold, particularly in the low-gain channels where the energy threshold is highest. This is the mechanism responsible for the residual gap between the geometric acceptance (71\%) and the measured efficiency (35\%). It illustrates how lowering the per-pixel noise, for example through cooling of the CMOS sensors or averaging over multiple readouts, would recover detection efficiency.

The threshold effect is confirmed directly by the channel composition of the cosmic ray coincidences. Accounting for the measured atmospheric muon flux and angular distribution, we predict the muon rate through each combination of trap-side and scintillator-side gain modes: high-gain (HG) and low-gain (LG). Because the HG–HG and LG–LG combinations require diagonal trajectories, we expect 9.1 events each, whereas the vertically stacked HG–LG and LG–HG combinations yield 13.6 events each. Comparing to the observed counts gives a per-combination efficiency. The HG–HG combination is fully recovered, with 12 observed events consistent with the prediction, whereas the LG–LG combination—requiring an above-threshold deposit in a low-gain channel on both planes—yields only 2 events, an efficiency of just $22\pm7\%$. This ordering, with HG–HG seeing far more events than LG–LG relative to expectation, confirms that the low-gain channels suffer from a higher effective threshold and thus a lower detection probability for the few-keV deposits characteristic of minimally ionizing particles. Despite the limited statistics, this directly demonstrates that the low-gain channels dominate the efficiency loss and underscores that reducing per-pixel noise will improve detection efficiency.

\subsection{Angular Resolution}

\begin{figure}[!t]
	\centering
    \includegraphics[width=\linewidth]{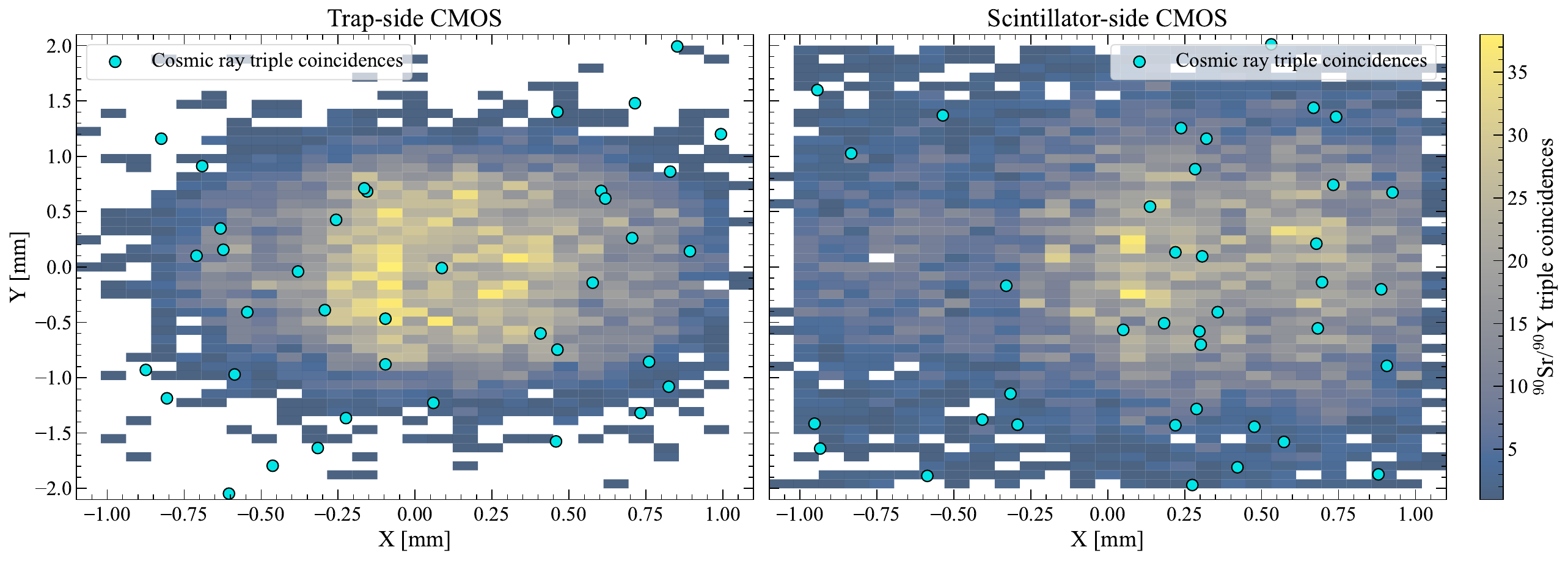}
	\caption{The distribution of triple coincidence events in each CMOS detector for $^{90}$Sr/$^{90}$Y (histogram) and cosmic rays (cyan dots). The spot size on the trap-side CMOS corresponds to the aperture of the collimated source of electrons. }
	\label{fig:TC_events}
\end{figure}

\begin{figure}[!t]
    \centering
    \includegraphics[width=0.48\linewidth]{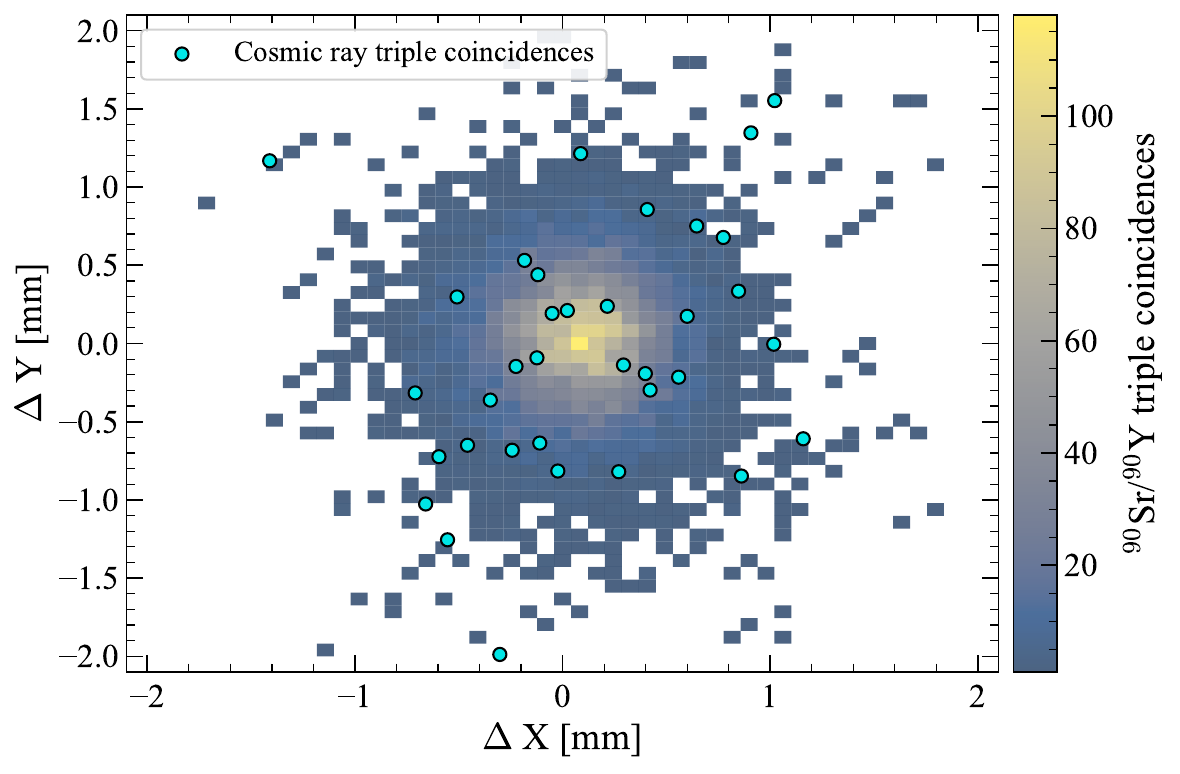}
    \includegraphics[width=0.48\linewidth]{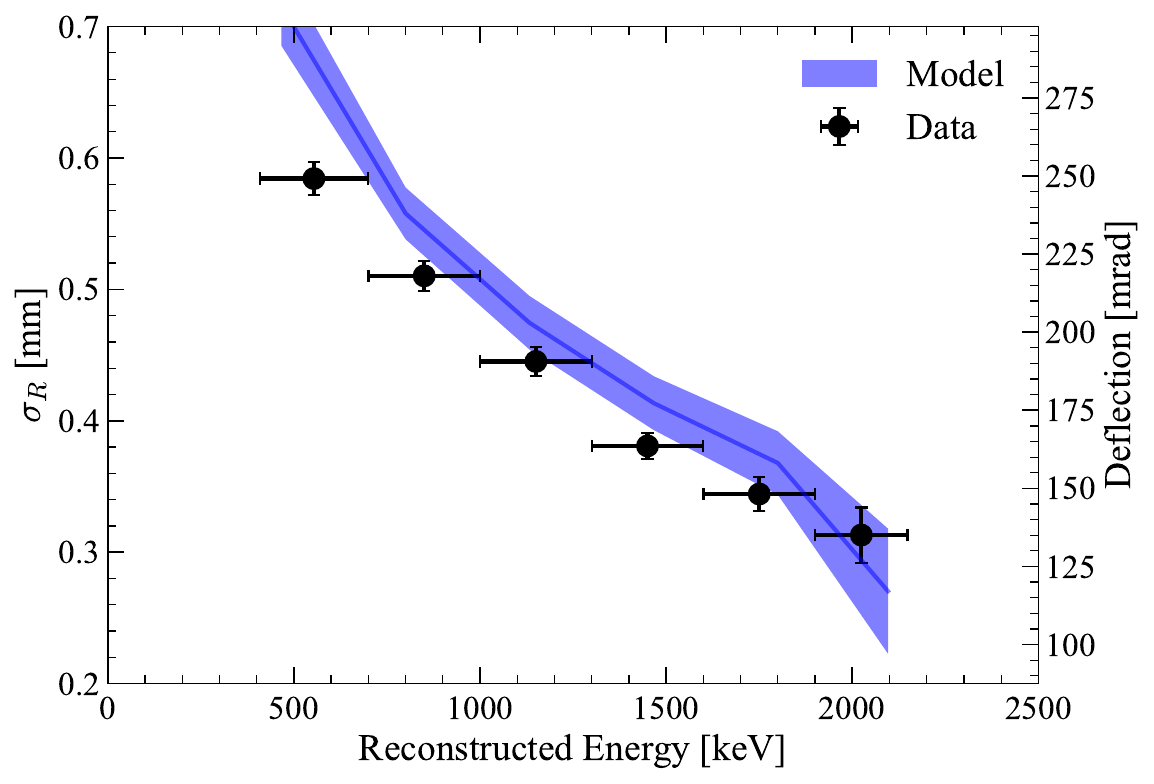}
    \caption{\emph{Left:} The ($\Delta$X,$\Delta$Y) of all reconstructed triple coincidences are shown for $^{90}$Sr/$^{90}$Y data (histogram) and cosmic rays (cyan dots). \emph{Right:} The radial spread of events between the two CMOS as a function of reconstructed scintillator energy. The data points (black) are compared to predictions from the dedicated \textsc{Geant4} model. The amount of deflection due to multiple scattering decreases with increasing electron energy.}
    \label{fig:TC_scatter}
\end{figure}

The distribution of the triple coincidence events is shown in Fig.~\ref{fig:TC_events}. The extent of the collimated source is evident in the spot size on the trap-side CMOS. The distribution of events on the scintillator-side CMOS is smeared due to the multiple scattering occurring in the silicon before reaching its active area. The central value of the distribution on the trap-side CMOS can also be seen to be shifted by about 200~$\mu$m in the +x-direction. The preference for events in the +x-direction on the scintillator-side CMOS is due to detection bias from the high gain channels (with $x>0$) and potentially slight misalignment in the CMOS attached to the board. 

The inter-CMOS hit displacement ($\Delta$X,$\Delta$Y) between the trap-side and scintillator-side CMOS is shown in the left plot in Fig.~\ref{fig:TC_scatter}. This displacement provides a direct measure of the multiple-Coulomb scattering of electrons as a function of their reconstructed kinetic energy. This scattering is determined by the amount of silicon between the two active epitaxial layers which are separated by 2.27~mm of which 22$\mu$m is silicon from the BEOL and silicon dioxide layer of each CMOS. The distance is set by the thickness of the CMOS board (1.71~mm) and the silicon frame mounting the CMOS on the board.

The extent to which multiple scattering occurs between the CMOS is displayed as the black data points in the right plot of Fig.~\ref{fig:TC_scatter}. We quantify the multiple scattering by calculating the RMS radius, $\sigma_{R}$, as the quadrature sum of $\sigma_{X}$ and $\sigma_{Y}$. These are measured by performing Gaussian fits to the $\Delta X$ and $\Delta Y$ as a function of reconstructed scintillator energy. The fitted $\sigma_{R}$ decreases monotonically with increasing energy as expected. The RMS spread is converted into a deflection angle as the arctangent of the ratio of $\sigma_{R}$ and the CMOS separation. The blue bands correspond to the same value as predicted by \textsc{Geant4} simulations, which is in reasonable agreement with the data.

It is worth clarifying the distinct roles of the two CMOS planes, since the large deflections in Fig.~\ref{fig:TC_scatter} might appear to compromise the momentum reconstruction. The inter-plane displacement corresponds to approximately 125–250~mrad of multiple scattering in the intervening silicon, which degrades the two-plane vertexing used to extrapolate the trajectory back to its origin. The $\beta$ direction, however, is defined by the hit on the trap-side CMOS—the plane encountered first, before this scattering occurs—together with the point-like emission source. The vertexing instead serves only to accept or reject events based on whether they point back to the source. The two are therefore decoupled, and the scattering that limits vertexing does not enter the momentum reconstruction.

\subsection{Pointing Resolution}

The excellent angular resolution described above enables the QuIPS $\beta$ momentum detector to perform vertexing of the source of electrons on an event-by-event basis. In this calibration geometry, this capability is assessed by extrapolating the straight-line trajectory defined by the cluster location in each CMOS to the plane of the collimator aperture. The collimator aperture has a diameter of 1.5~mm, which defines the geometric extent of the true source distribution.

\begin{figure}[!t]
    \centering
    \includegraphics[width=0.48\linewidth]{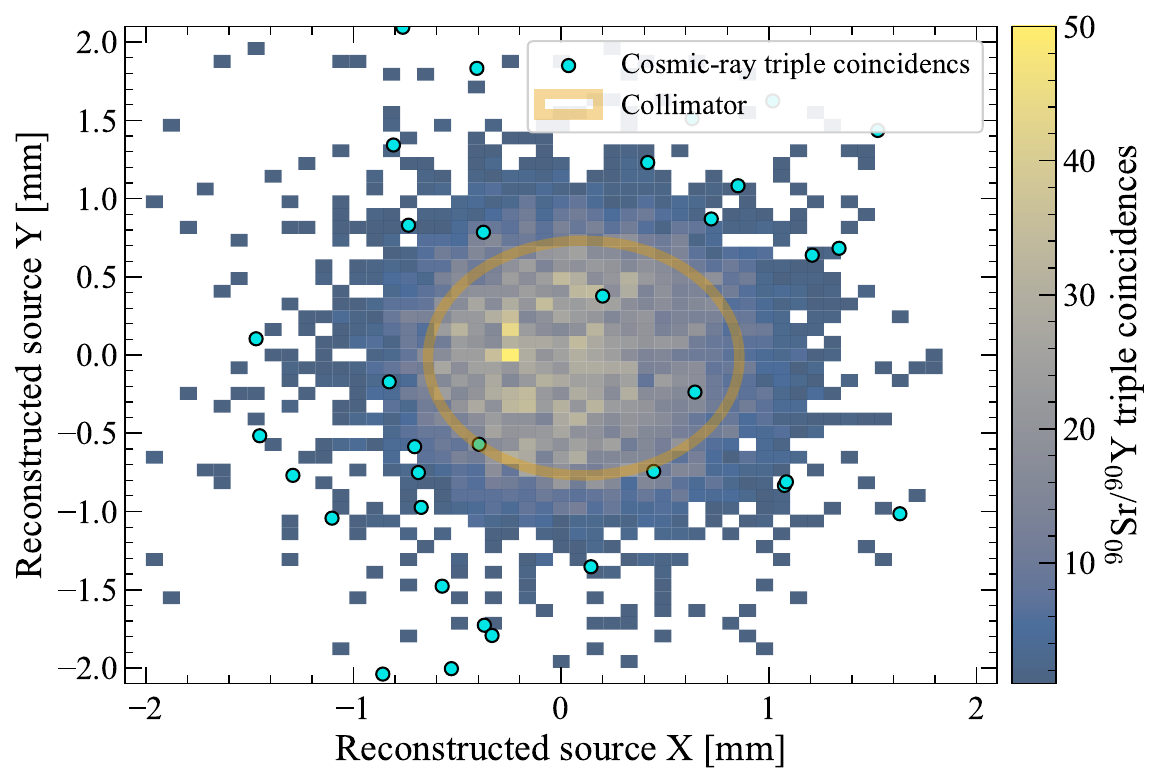}
    \includegraphics[width=0.48\linewidth]{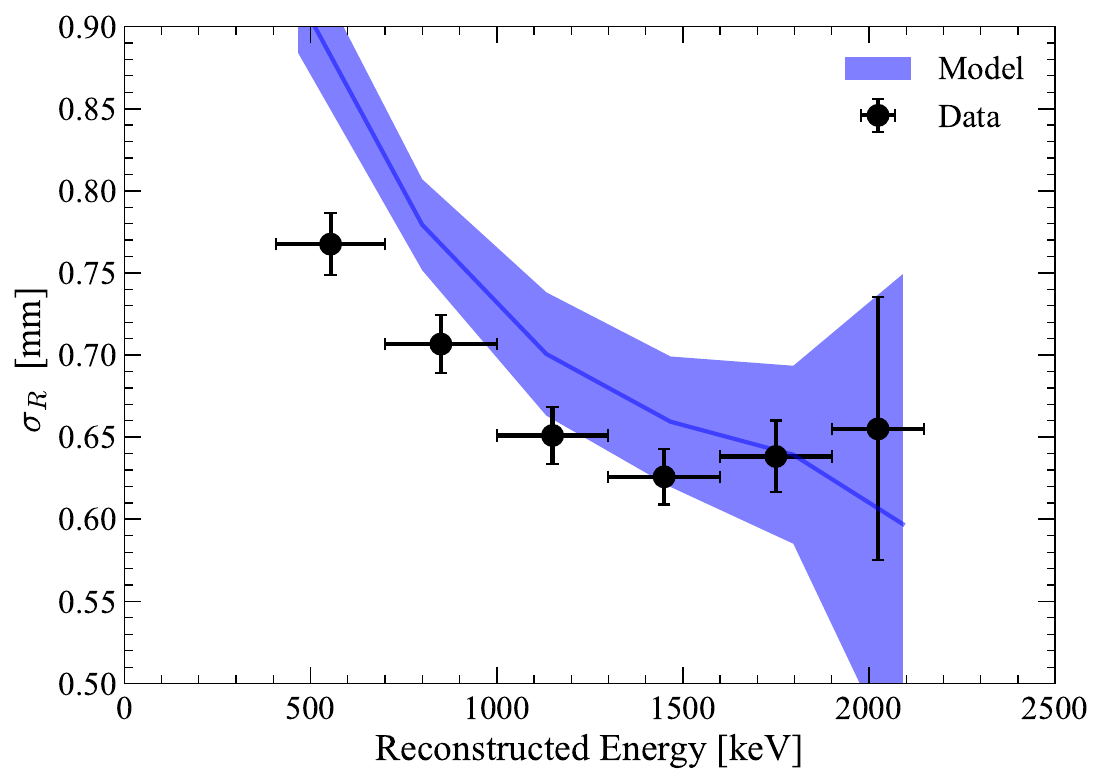}
    \caption{\emph{Left:} Reconstructed location of the source of collimated electrons produced by extrapolating a straight line through the electron hits in each CMOS to the plane of the collimator aperture. \emph{Right:} The RMS radius of the reconstructed source location as a function of energy (black) is compared to predictions from the dedicated \textsc{Geant4} model.}
    \label{fig:SourceRecon}
\end{figure}

The left plot in Fig.~\ref{fig:SourceRecon} shows reconstructed source positions for all $^{90}$Sr/$^{90}$Y triple coincidences with the 1.5~mm collimator aperture overlaid. The triple coincidence cosmic rays are depicted as cyan dots. Critically, these events show no fixed pointing to any source as expected. The right plot of Fig.~\ref{fig:SourceRecon} shows the reconstructed RMS radius of the source as a function of reconstructed energy, calculated identically to the inter-CMOS distribution described in the previous section. The data agree well with predictions from the \textsc{Geant4} model, shown as the blue band.

The fitted radial extent of the reconstructed source distribution decreases systematically with increasing $\beta$ energy, from approximately 0.77~mm at scintillator energies of 500~keV to approximately 0.63~mm at 1500~keV. At low energies, the reconstructed source radius is comparable to or slightly larger than the collimator radius, as expected because multiple Coulomb scattering in the first CMOS layer and in the intervening dead material causes a spread in the reconstructed emission point that adds in quadrature to the geometric collimator size. At higher energies, the multiple-scattering contribution decreases and the reconstructed source radius converges toward the geometric collimator size. 

Importantly, the pointing resolution is sufficient to reject events that do not originate from the collimator. Cosmic ray muons, which traverse the detector from above and are minimally ionizing, will have minimal deflections between the two CMOS. However, their reconstructed source position should not point toward any fixed location in the collimator plane as confirmed in Fig.~\ref{fig:SourceRecon}. This confirms that the pointing reconstruction correctly discriminates between isotropically incident cosmic rays and collimated $\beta$ particles originating from a fixed source location, validating that this detector will be able to correctly identify events originating from a levitated nanosphere. Even more critically, electrons that scatter off the lenses used to form the optical trap and hit the detectors can be rejected since the lens surfaces are displaced from the trap by distances exceeding the pointing resolution.

\subsection{Discussion of Systematic Uncertainties}

Several systematic effects relevant to the combined calibration are identified and discussed below. We emphasize that a number of these are specific to the calibration configuration, in particular the use of a collimated external source, and do not apply to the eventual deployment of the detector with an optically levitated nanosphere.

The dominant systematic uncertainty arises from the alignment of the radioactive source with the collimator aperture. The $^{90}$Sr/$^{90}$Y source is a finite-extent deposited source, approximately 2.5~mm in diameter, while the collimator aperture is only 1.5~mm in diameter. Further, the source was manually placed beneath the collimator and may have been imperfectly aligned. Electrons emitted from regions of the source that are not directly beneath the aperture must scatter off the walls of the copper collimator before reaching the detector, losing energy in the process. This degrades the reconstructed energy and, in particular, suppresses the high-energy end of the spectrum, shifting the observed endpoint downward. The shaded bands in Fig.~\ref{fig:eff} depict the bounding cases of when the electron is emitted directly under the collimator or outside its radial extent. Comparison with the data indicates that the observed spectrum prefers a population that is, on average, slightly more misaligned than perfect alignment with the aperture would imply. 

Critically, this alignment systematic is an artifact of the collimated-source calibration geometry and is not a concern for the full QuIPS setup. In the deployed configuration, the $\beta$-emitting radioisotope is embedded within an optically levitated nanosphere of order 100~nm in diameter, which constitutes a genuinely point-like and well-localized source with no surrounding collimator off which electrons can scatter. The energy-loss mechanism that drives this systematic in the calibration is therefore absent in the physics configuration, and the reconstructed $\beta$ energy spectrum in deployment will not be subject to the same endpoint degradation.

The next most significant systematic arises from the incomplete modeling of the CMOS sensor response in the \textsc{Geant4} simulation. The simulation treats the active epitaxial layer as a uniform pixelated silicon volume and does not capture the full dynamics of charge diffusion and collection in the sensor, nor charge sharing during the readout. This is evident in the comparison between data and simulation. The observed pixel clusters are systematically larger than those predicted by the model, and the fraction of energy deposited in neighboring pixels is correspondingly larger in data, consistent with charge spreading from diffusion and from the readout that is not present in the simulation. Because the detection threshold is applied per pixel, this mismodeling directly affects the predicted efficiency and the reconstructed CMOS energy spectrum, and it is the leading explanation for the difference between the simulated efficiency of 58\% and the measured value of 35\%. A related question is whether the full 10~$\mu$m thickness of the epitaxial layer is active in the data. The simulation assumes that it is, and any difference would further shift the effective per-pixel energy scale relative to the model.

Additional, subdominant systematics are also identified. The energy calibration of the scintillator-SiPM system was performed at a different bias voltage than the full-stack data. We correct for voltage-dependent effects in the SiPM response. We expect any residual uncertainty, largely manifesting in the observed endpoint, to be minimal. On the CMOS side, the thinning process left approximately 20-30~$\mu$m of additional silicon per chip beyond the nominal layers, which is included in the simulation as dead material but not modeled with its full spatial profile; this contributes a small uncertainty to the multiple-scattering and energy-loss predictions, consistent with the mild disagreement observed between the data and the simulated angular deflection. Finally, bremsstrahlung photons produced in the detector materials can deposit energy in the scintillator and be convolved with genuine $\beta$ events; the simulation indicates that this contribution is concentrated at low energies and is subdominant to genuine $\beta$ particles across the analysis range above 500~keV.

%%%%%%%%%%%%%%%%%%%%%%%%%%%%%%%%%%%%%%%%%%%%%%%%%%%
\section{Deployment and Outlook}
\label{sec:prospects}

We have presented the design, development, and first calibration results of a compact electron detector for the QuIPS experiment. The detector combines two thinned CMOS sensors for directional $\beta$-particle tracking with an EJ-200 plastic scintillator read out by Hamamatsu S14161-6050HS silicon photomultipliers for calorimetry, all operating in ultra-high vacuum inside an optical trapping chamber. 

The full-system calibration demonstrates that the detector meets the core requirements of the QuIPS physics program. We detect $\beta$ particles with an overall efficiency of approximately 35\% from 500~keV to the $^{90}$Y endpoint, the calorimeter achieves an energy resolution of 5\% at 1~MeV with a linear response to the endpoint, and the CMOS tracker reconstructs the $\beta$ emission point to better than 1~mm. This pointing capability cleanly discriminates collimated $\beta$ particles from isotropically incident cosmic rays, validating the strategy for rejecting backgrounds in the deployed experiment.

The detector is now being prepared for deployment in the nanosphere optical levitation chamber at Yale University (see description in Ref.~\cite{Tseng:2025rlo}), where it will be installed above the optical trap and certified to operate in concert with the levitated optomechanical setup. This will constitute, to our knowledge, the first demonstration of a particle detector operating in unison with an optically levitated nanosphere. Several detector improvements have also been identified: reading out the full set of potentially active CMOS pixels, cooling the SiPM and CMOS, and averaging over multiple readouts to reduce noise, improving the thinning uniformity, and refining the \textsc{Geant4} modeling. In parallel, a dedicated experiment is planned at LBNL to operate a large array of levitated sensors with impulse sensitivity at the standard quantum limit. This work represents the first development of a UHV-compatible, integrated electron tracking and calorimetry detector designed for operation with a quantum optomechanical sensor, and opens a new pathway for precision measurements in neutrino physics and beyond-Standard-Model searches.

%\appendix
%\section{Appendix 1}

\acknowledgments

We thank Ben Knepper, Giacomo Marocco, Dave Rauchwerk, and Emil Rofors for early discussion and collaboration, Nowzesh Hasan and Daniel Bondi for help with machining and fabrication, and the QuIPS team in the Moore Lab at Yale for their continued collaboration. This material is based upon work supported by the Laboratory Directed Research and Development Program of Lawrence Berkeley National Laboratory under U.S. Department of Energy Contract No. DE-AC02-05CH11231, by U.S. Department of Energy Early Career Research Award DE-SCL0000025, and by the Gordon and Betty Moore Foundation's Mid-Career Experimental Investigator Award GBMF13781.

%\paragraph{Note added.} 

% Bibliography

%% [A] Recommended: using JHEP.bst file
%% \bibliographystyle{JHEP}
%% \bibliography{biblio.bib}

%% or
%% [B] Manual formatting (see below)
%% (i) We suggest to always provide author, title and journal data or doi:
%% in short all the informations that clearly identify a document.
%% (ii) please avoid comments such as "For a review'', "For some examples",
%% "and references therein" or move them in the text. In general, please leave only references in the bibliography and move all
%% accessory text in footnotes.
%% (iii) Also, please have only one work for each \bibitem.

\bibliographystyle{JHEP}
\bibliography{biblio.bib}

@article{Carney:2022pku,
    author = "Carney, Daniel and Leach, Kyle G. and Moore, David C.",
    title = "{Searches for Massive Neutrinos with Mechanical Quantum Sensors}",
    eprint = "2207.05883",
    archivePrefix = "arXiv",
    primaryClass = "hep-ex",
    doi = "10.1103/PRXQuantum.4.010315",
    journal = "PRX Quantum",
    volume = "4",
    number = "1",
    pages = "010315",
    year = "2023"
}

@article{smith2019proposed,
  title={Proposed experiments to detect keV-range sterile neutrinos using energy-momentum reconstruction of beta decay or K-capture events},
  author={Smith, Peter F},
  journal={New Journal of Physics},
  volume={21},
  number={5},
  pages={053022},
  year={2019},
  publisher={IOP Publishing}
}

@article{martoff2021hunter,
  title={HUNTER: precision massive-neutrino search based on a laser cooled atomic source},
  author={Martoff, CJ and others},
  journal={Quantum Science \& Technology},
  volume={6},
  number={2},
  pages={024008},
  year={2021},
  publisher={IOP Publishing}
}

@article{KATRIN:2024cdt,
    author = "Aker, Max and others",
    collaboration = "KATRIN",
    title = "{Direct neutrino-mass measurement based on 259 days of KATRIN data}",
    eprint = "2406.13516",
    archivePrefix = "arXiv",
    primaryClass = "nucl-ex",
    doi = "10.1126/science.adq9592",
    journal = "Science",
    volume = "388",
    number = "6743",
    pages = "adq9592",
    year = "2025"
}

@article{Project8:2017nal,
    author = "Ashtari Esfahani, Ali and others",
    collaboration = "Project 8",
    title = "{Determining the neutrino mass with cyclotron radiation emission spectroscopy{\textemdash}Project 8}",
    eprint = "1703.02037",
    archivePrefix = "arXiv",
    primaryClass = "physics.ins-det",
    doi = "10.1088/1361-6471/aa5b4f",
    journal = "J. Phys. G",
    volume = "44",
    number = "5",
    pages = "054004",
    year = "2017"
}

@article{ACCESS:2023gdy,
    author = "Pagnanini, L. and others",
    collaboration = "ACCESS",
    title = "{Array of cryogenic calorimeters to evaluate the spectral shape of forbidden $\beta $-decays: the ACCESS project}",
    eprint = "2305.01949",
    archivePrefix = "arXiv",
    primaryClass = "physics.ins-det",
    doi = "10.1140/epjp/s13360-023-03946-x",
    journal = "Eur. Phys. J. Plus",
    volume = "138",
    number = "5",
    pages = "445",
    year = "2023"
}

@article{ANDREOTTI2007208,
title = {MARE, Microcalorimeter Arrays for a Rhenium Experiment: A detector overview},
journal = {Nuclear Instruments and Methods in Physics Research Section A: Accelerators, Spectrometers, Detectors and Associated Equipment},
volume = {572},
number = {1},
pages = {208-210},
year = {2007},
note = {Frontier Detectors for Frontier Physics},
issn = {0168-9002},
doi = {https://doi.org/10.1016/j.nima.2006.10.198},
url = {https://www.sciencedirect.com/science/article/pii/S016890020602064X},
author = {E. Andreotti and others},
}

@article{Nab:2018toa,
    author = "Fry, J. and others",
    editor = "Jenke, T. and Degenkolb, S. and Geltenbort, P. and Jentschel, M. and Nesvizhevsky, V. V. and Rebreyend, D. and Roccia, S. and Soldner, T. and Stutz, A. and Zimmer, O.",
    collaboration = "Nab",
    title = "{The Nab Experiment: A Precision Measurement of Unpolarized Neutron Beta Decay}",
    eprint = "1811.10047",
    archivePrefix = "arXiv",
    primaryClass = "nucl-ex",
    doi = "10.1051/epjconf/201921904002",
    journal = "EPJ Web Conf.",
    volume = "219",
    pages = "04002",
    year = "2019"
}

@article{Hassan:2020hrj,
    author = "Hassan, M. T. and others",
    title = "{Measurement of the neutron decay electron-antineutrino angular correlation by the aCORN experiment}",
    eprint = "2012.14379",
    archivePrefix = "arXiv",
    primaryClass = "nucl-ex",
    doi = "10.1103/PhysRevC.103.045502",
    journal = "Phys. Rev. C",
    volume = "103",
    number = "4",
    pages = "045502",
    year = "2021"
}

@article{NEXT:2023daz,
    author = "Novella, P. and others",
    collaboration = "NEXT",
    title = "{Demonstration of neutrinoless double beta decay searches in gaseous xenon with NEXT}",
    eprint = "2305.09435",
    archivePrefix = "arXiv",
    primaryClass = "nucl-ex",
    reportNumber = "FERMILAB-PUB-23-251-ND",
    doi = "10.1007/JHEP09(2023)190",
    journal = "JHEP",
    volume = "09",
    pages = "190",
    year = "2023"
}

@article{SuperNEMO:2010wnd,
    author = "Arnold, R. and others",
    collaboration = "SuperNEMO",
    title = "{Probing New Physics Models of Neutrinoless Double Beta Decay with SuperNEMO}",
    eprint = "1005.1241",
    archivePrefix = "arXiv",
    primaryClass = "hep-ex",
    reportNumber = "MAN-HEP-2010-2",
    doi = "10.1140/epjc/s10052-010-1481-5",
    journal = "Eur. Phys. J. C",
    volume = "70",
    pages = "927--943",
    year = "2010"
}

@article{Delahaye:2018nok,
    author = "Delahaye, P. and others",
    title = "{The open LPC Paul trap for precision measurements in beta decay}",
    eprint = "1810.09246",
    archivePrefix = "arXiv",
    primaryClass = "physics.ins-det",
    doi = "10.1140/epja/i2019-12777-3",
    journal = "Eur. Phys. J. A",
    volume = "55",
    number = "6",
    pages = "101",
    year = "2019"
}

@article{SHIDLING2021116636,
title = "The TAMUTRAP facility: A Penning trap facility at Texas A\&M University for weak interaction studies",
journal = {International Journal of Mass Spectrometry},
volume = {468},
pages = {116636},
year = {2021},
issn = {1387-3806},
doi = {https://doi.org/10.1016/j.ijms.2021.116636},
url = {https://www.sciencedirect.com/science/article/pii/S1387380621001160},
author = {P.D. Shidling and others}
}

@article{Burkey:2022gpb,
    author = "Burkey, M. T. and others",
    title = "{Improved Limit on Tensor Currents in the Weak Interaction from Li8 {\ensuremath{\beta}} Decay}",
    eprint = "2205.01865",
    archivePrefix = "arXiv",
    primaryClass = "nucl-ex",
    doi = "10.1103/PhysRevLett.128.202502",
    journal = "Phys. Rev. Lett.",
    volume = "128",
    number = "20",
    pages = "202502",
    year = "2022"
}

@article{Vetter:2008zz,
    author = "Vetter, P. A. and Abo-Shaeer, J. R. and Freedman, S. J. and Maruyama, R.",
    title = "{Measurement of the beta-nu correlation of Na-21 using shakeoff electrons}",
    eprint = "0805.1212",
    archivePrefix = "arXiv",
    primaryClass = "nucl-ex",
    doi = "10.1103/PhysRevC.77.035502",
    journal = "Phys. Rev. C",
    volume = "77",
    pages = "035502",
    year = "2008"
}

@article{Muller:2022jew,
    author = {M{\"u}ller, P. and others},
    title = "{$\beta$-nuclear-recoil correlation from $^6$He decay in a laser trap}",
    eprint = "2206.00742",
    archivePrefix = "arXiv",
    primaryClass = "nucl-ex",
    doi = "10.1103/PhysRevLett.129.182502",
    journal = "Phys. Rev. Lett.",
    volume = "129",
    number = "18",
    pages = "182502",
    year = "2022"
}

@article{Delic:2019xqd,
    author = "Deli{\'c}, Uro{\v{s}} and Reisenbauer, Manuel and Dare, Kahan and Grass, David and Vuleti{\'c}, Vladan and Kiesel, Nikolai and Aspelmeyer, Markus",
    title = "{Cooling of a levitated nanoparticle to the motional quantum ground state}",
    eprint = "1911.04406",
    archivePrefix = "arXiv",
    primaryClass = "quant-ph",
    doi = "10.1126/science.aba3993",
    journal = "Science",
    volume = "367",
    number = "6480",
    pages = "892--895",
    year = "2020"
}

@article{Tebbenjohanns:2021tgw,
    author = "Tebbenjohanns, Felix and Mattana, M. Luisa and Rossi, Massimiliano and Frimmer, Martin and Novotny, Lukas",
    title = "{Quantum control of a nanoparticle optically levitated in cryogenic free space}",
    eprint = "2103.03853",
    archivePrefix = "arXiv",
    primaryClass = "quant-ph",
    doi = "10.1038/s41586-021-03617-w",
    journal = "Nature",
    volume = "595",
    number = "7867",
    pages = "378--382",
    year = "2021"
}

@article{Magrini2021,
       author = {{Magrini}, Lorenzo and {Rosenzweig}, Philipp and {Bach}, Constanze and {Deutschmann-Olek}, Andreas and {Hofer}, Sebastian G. and {Hong}, Sungkun and {Kiesel}, Nikolai and {Kugi}, Andreas and {Aspelmeyer}, Markus},
        title = "{Real-time optimal quantum control of mechanical motion at room temperature}",
      journal = {Nature},
         year = 2021,
        month = jul,
       volume = {595},
       number = {7867},
        pages = {373-377},
          doi = {10.1038/s41586-021-03602-3},
archivePrefix = {arXiv},
       eprint = {2012.15188},
 primaryClass = {quant-ph},
       adsurl = {https://ui.adsabs.harvard.edu/abs/2021Natur.595..373M}
}

@article{Ranjit:2016dlq,
    author = "Ranjit, Gambhir and Cunningham, Mark and Casey, Kirsten and Geraci, Andrew A.",
    title = "{Zeptonewton force sensing with nanospheres in an optical lattice}",
    eprint = "1603.02122",
    archivePrefix = "arXiv",
    primaryClass = "physics.optics",
    doi = "10.1103/PhysRevA.93.053801",
    journal = "Phys. Rev. A",
    volume = "93",
    number = "5",
    pages = "053801",
    year = "2016"
}

@article{Monteiro:2020qiz,
    author = "Monteiro, Fernando and Li, Wenqiang and Afek, Gadi and Li, Chang-ling and Mossman, Michael and Moore, David C.",
    title = "{Force and acceleration sensing with optically levitated nanogram masses at microkelvin temperatures}",
    eprint = "2001.10931",
    archivePrefix = "arXiv",
    primaryClass = "physics.optics",
    doi = "10.1103/PhysRevA.101.053835",
    journal = "Phys. Rev. A",
    volume = "101",
    number = "5",
    pages = "053835",
    year = "2020"
}

@article{Tseng:2025rlo,
    author = "Tseng, Yu-Han and Penny, T. W. and Siegel, Benjamin and Wang, Jiaxiang and Moore, David C.",
    title = "{Search for Dark Matter Scattering from Optically Levitated Nanoparticles}",
    eprint = "2508.00815",
    archivePrefix = "arXiv",
    primaryClass = "hep-ex",
    doi = "10.1103/j76m-gcp1",
    journal = "PRX Quantum",
    volume = "6",
    number = "4",
    pages = "040367",
    year = "2025"
}

@article{Tseng:2026flh,
    author = "Tseng, Yu-Han and Hardy, Clarke A. and Penny, T. W. and Lowe, Cecily and Baeza-Rubio, Jacqueline and Carney, Daniel and Moore, David C.",
    title = "{Optomechanical Detection of Individual Gas Collisions}",
    eprint = "2604.18371",
    archivePrefix = "arXiv",
    primaryClass = "quant-ph",
    month = "4",
    year = "2026"
}

@article{WANG20181,
title = {The impact of incorporating shell-corrections to energy loss in silicon},
journal = {Nuclear Instruments and Methods in Physics Research Section A: Accelerators, Spectrometers, Detectors and Associated Equipment},
volume = {899},
pages = {1-5},
year = {2018},
issn = {0168-9002},
doi = {https://doi.org/10.1016/j.nima.2018.04.061},
url = {https://www.sciencedirect.com/science/article/pii/S0168900218305783},
author = {Fuyue Wang and Su Dong and Benjamin Nachman and Maurice Garcia-Sciveres and Qi Zeng}
}

@inproceedings{Lee1995AnAP,
  title={An Active Pixel Sensor Fabricated Using CMOS / CCD Process Technology},
  author={Paul P. K. Lee and Russell C. Gee and R. Michael Guidash and T.-H. Lee and Eric R. Fossum},
  year={1995},
  booktitle="",
}

@article{PhysRevLett.56.2195,
  title = {Monovacancy Formation Enthalpy in Silicon},
  author = {Dannefaer, S. and Mascher, P. and Kerr, D.},
  journal = {Phys. Rev. Lett.},
  volume = {56},
  issue = {20},
  pages = {2195--2198},
  numpages = {0},
  year = {1986},
  month = {May},
  publisher = {American Physical Society},
  doi = {10.1103/PhysRevLett.56.2195},
  url = {https://link.aps.org/doi/10.1103/PhysRevLett.56.2195}
}

@manual{hamamatsu_S14161,
  title        = {MPPC (Multi-Pixel Photon Counter) S14160/14161 series datasheets},
  organization = {Hamamatsu Photonics},
  address      = {Hamamatsu City, Japan},
  year         = {2026},
  url          = {https://www.hamamatsu.com/content/dam/hamamatsu-photonics/sites/documents/99_SALES_LIBRARY/ssd/s14160_s14161_series_kapd1064e.pdf},
}

@article{NepomukOtte:2016ktf,
    author = "Nepomuk Otte, Adam and Garcia, Distefano and Nguyen, Thanh and Purushotham, Dhruv",
    title = "{Characterization of Three High Efficiency and Blue Sensitive Silicon Photomultipliers}",
    eprint = "1606.05186",
    archivePrefix = "arXiv",
    primaryClass = "physics.ins-det",
    doi = "10.1016/j.nima.2016.09.053",
    journal = "Nucl. Instrum. Meth. A",
    volume = "846",
    pages = "106--125",
    year = "2017"
}

@manual{Eljen_EJ200,
  title        = {General purpose plastic scintillator EJ-200, EJ-204, EJ-208, EJ-212},
  organization = {Eljen Technology},
  address      = {Sweetwater, Texas, USA},
  year         = {2026},
  url          = {https://eljentechnology.com/images/products/data_sheets/EJ-200_EJ-204_EJ-208_EJ-212.pdf},
}

@article{Nassalski_2008,
  author={Nassalski, A. and Moszynski, M. and Syntfeld-Kazuch, A. and Swiderski, L. and Szczeniak, T.},
  journal={IEEE Transactions on Nuclear Science}, 
  title={Non-Proportionality of Organic Scintillators and BGO}, 
  year={2008},
  volume={55},
  number={3},
  pages={1069-1072},
  doi={10.1109/TNS.2007.913478}
}

@article{Payne_2011,
  author={Payne, Stephen A. and others},
  journal={IEEE Transactions on Nuclear Science}, 
  title={Nonproportionality of Scintillator Detectors: Theory and Experiment. II}, 
  year={2011},
  volume={58},
  number={6},
  pages={3392-3402},
  doi={10.1109/TNS.2011.2167687}
}

@article{Safari:2016ypc,
    author = "Safari, M. J. and Davani, F. Abbasi and Afarideh, H.",
    title = "{Differentiation method for localization of Compton edge in organic scintillation detectors}",
    eprint = "1610.09185",
    archivePrefix = "arXiv",
    primaryClass = "physics.ins-det",
    doi = "10.22034/rpe.2020.104839",
    month = "10",
    year = "2016",
    journal = ""
}

@article{TAJUDIN2020109086,
    title = {Response of plastic scintillator to gamma sources},
    journal = {Applied Radiation and Isotopes},
    volume = {159},
    pages = {109086},
    year = {2020},
    issn = {0969-8043},
    doi = {https://doi.org/10.1016/j.apradiso.2020.109086},
    url = {https://www.sciencedirect.com/science/article/pii/S0969804319301034},
    author = {Suffian M. Tajudin and Y. Namito and T. Sanami and H. Hirayama},
}

@article{CRAUN1970239,
    title = {Analysis of response data for several organic scintillators},
    journal = {Nuclear Instruments and Methods},
    volume = {80},
    number = {2},
    pages = {239-244},
    year = {1970},
    issn = {0029-554X},
    doi = {https://doi.org/10.1016/0029-554X(70)90768-8},
    url = {https://www.sciencedirect.com/science/article/pii/0029554X70907688},
    author = {R.L. Craun and D.L. Smith},
}

@article{Tran2018,
  author={Tran, Ngan N. T. and Sasaki, Shinichi and Sanami, Toshiya and Kishimoto, Yuji and Shibamura, Eido},
  journal={IEEE Transactions on Nuclear Science}, 
  title={Scintillation Efficiency and Position Sensitivity for Radiation Events in Plastic Scintillators}, 
  year={2018},
  volume={65},
  number={8},
  pages={2178-2183},
  doi={10.1109/TNS.2018.2805703}
}

@article{LSwiderski_2012,
doi = {10.1088/1748-0221/7/06/P06011},
url = {https://doi.org/10.1088/1748-0221/7/06/P06011},
year = {2012},
month = {jun},
publisher = {},
volume = {7},
number = {06},
pages = {P06011},
author = {L Swiderski and R Marcinkowski and M Moszynski and W Czarnacki and M Szawlowski and T Szczesniak and G Pausch and C Plettner and K Roemer},
title = {Electron response of some low-Z scintillators in wide energy range},
journal = {Journal of Instrumentation}
}

\end{document}